# A Semantic Framework for Patient Digital Twins in Chronic Care

AMAL ELGAMMAL[1,2,3], BERND J. KRÄMER[1,4], MICHAEL P. PAPAZOGLOU[1,5,6], MIRA RAHEEM [1,2]

[1]*Scientific Academy for Service Technology e.V. (ServTech), Potsdam, 14467, Germany*

[2]*Faculty of Computers & Artificial Intelligence, Cairo University, Cairo, 12613 Egypt*

[3]*Faculty of Computing and Information Sciences, Egypt University of Informatics, Cairo, 19519, Egypt*

[4]*FernUniversität, Hagen, 58097, Germany*

[5]*University of Macquarie, Sydney, 2000, Australia*

[6]*University of New Wales, Sydney NSW 2033, Australia*

**Abstract**

Personalized chronic care requires the integration of multimodal health data to enable precise, adaptive, and preventive decision-making. Yet most current digital twin (DT) applications remain organ-specific or tied to isolated data types, lacking a unified and privacy-preserving foundation. This paper introduces the Patient Medical Digital Twin (PMDT), an ontology-driven in silico patient framework that integrates physiological, psychosocial, behavioral, and genomic information into a coherent, extensible model. Implemented in OWL 2.0, the PMDT ensures semantic interoperability, supports automated reasoning, and enables reuse across diverse clinical contexts. Its ontology is structured around modular Blueprints (patient, disease and diagnosis, treatment and follow-up, trajectories, safety, pathways, and adverse events), formalized through dedicated conceptual views. These were iteratively refined and validated through expert workshops, questionnaires, and a pilot study in the EU H2020 QUALITOP project with real-world immunotherapy patients. Evaluation confirmed ontology coverage, reasoning correctness, usability, and GDPR compliance. Results demonstrate the PMDT's ability to unify heterogeneous data, operationalize competency questions, and support descriptive, predictive, and prescriptive analytics in a federated, privacy-preserving manner. By bridging gaps in data fragmentation and semantic standardization, the PMDT provides a validated foundation for next-generation digital health ecosystems, transforming chronic care toward proactive, continuously optimized, and equitable management.

# 1 Introduction

Digital Twin (DT) technology enables the creation of dynamic digital representations of real-world entities that continuously evolve with their physical counterparts. In healthcare, DTs promise to transform clinical practice by enabling individualized monitoring, decision support, and proactive interventions, thereby improving patient outcomes and reducing costs [1]. Chronic disease management is a particularly compelling domain: conditions such as diabetes, cardiovascular disease, and cancer account for nearly three-quarters of global deaths [2], yet many are preventable through early detection, targeted interventions, and continuous monitoring. By integrating longitudinal, multimodal data, patient-specific digital twins can shift chronic care from episodic treatment toward continuous, adaptive, and preventive management.

Despite this potential, existing healthcare DTs remain fragmented. Most efforts are disease- or organ-specific, with limited ability to combine heterogeneous data streams such as electronic health records (EHRs), patient-generated health data (PGHD), and genomics. Semantic standardization is often absent, preventing automated reasoning and reuse across contexts. Moreover, privacy-preserving and federated analytics, essential for adoption in multi-institutional settings, are rarely incorporated. As a result, current DT solutions fall short of delivering a comprehensive, patient-centric view that is interoperable, extensible, and clinically usable.

**Research Gap.** There is currently no comprehensive patient digital twin framework that:

1. Integrates multimodal health data spanning physiological, psychosocial, behavioral, and genomic determinants.
2. Employs formal semantics to ensure interoperability, automated reasoning, and extensibility.
3. Embeds privacy-preserving, federated analytics to enable collaboration across institutions.

**Contribution.** This paper addresses this gap by introducing the *Patient Medical Digital Twin (PMDT)*, a formally modeled, ontology-driven in silico patient framework. Implemented in OWL 2.0, the PMDT organizes multimodal evidence into modular *Blueprints* (patient, disease and diagnosis, treatment and follow-up, trajectories, safety, pathways, and adverse events), each capturing a critical dimension of the patient profile. The ontology enables semantic interoperability, consistency checking, and SPARQL-based reasoning, while aligning with external standards such as the Disease Ontology [3] and OWL-Time [4].

The PMDT was developed iteratively within the EU H2020 *QUALITOP project*, where its conceptual design and ontology implementation were validated through expert workshops, questionnaires, and a multi-country pilot study involving immunotherapy patients. This grounding ensured that modeling decisions directly addressed real-world requirements, from longitudinal monitoring of immune-related adverse events (irAEs) to the integration of quality-of-life (QoL) measures and lifestyle data.

**Research Question**: *How can a patient-centric digital twin be formally modeled as an ontology to (i) integrate multimodal health data, (ii) ensure semantic interoperability and automated reasoning, and (iii) enable privacy-preserving, federated analytics for personalized chronic care?*

**Scope**: While the PMDT provides a foundation for advanced descriptive, predictive, and prescriptive analytics, the present paper focuses on its conceptual and modeling contributions: the ontology design, its modular Blueprint-based structure, and its integration within a federated architecture. Evaluation of predictive models and analytics pipelines is ongoing and will be reported separately.

**Structure of the paper:** Section 2 reviews related work on healthcare digital twins and ontology-driven health data integration. Section 3 introduces the PMDT vision and requirements, grounded in clinical personas and pilot scenarios. Section 4 details the ontology design and conceptual views, while Section 5 presents its implementation in OWL and integration into a federated ecosystem. Section 6 reports on validation and evaluation in the QUALITOP pilot study. Section 7 concludes with future directions, including blockchain-enabled data governance and expanded clinical evaluation.

## 2 Related Work

Digital Twin (DT) technology in medicine remains in its early stages and is often described as the convergence of data science, software engineering, expert knowledge representation, and AI/machine learning. These components underpin tools that support effective clinical decision-making and predictive modelling.

Vallée [5] provides a comprehensive review of digital twin applications in healthcare, acknowledging their *potential* to integrate heterogeneous data sources (EHRs, wearable sensors, IoT devices, and omics data) for personalized patient profiling, predictive modelling, and preventive interventions. The review further discusses envisioned benefits such as optimizing clinical operations and enabling risk-free training and simulation. At the same time, Vallée highlights that these capabilities remain largely aspirational: persistent challenges around interoperability, data privacy, and clinical validation still limit the realization of such applications in practice.

A scoping review [6] identified four essential DT characteristics: continuous updates, high fidelity to real-world conditions, decision-making feedback loops, and multifunctionality. Most reported studies lacked at least one of these characteristics, with multifunctionality being especially rare. A systematic review [1] of emerging healthcare data modelling technologies, including eight DT-related studies [7-14], highlighted narrow analytical scopes, limited integration of heterogeneous data, and persistent issues with interoperability, usability, scalability, and validation. A meta-review [15] echoed these findings, pointing to technical barriers, scalability issues, and insufficient clinical validation as key adoption hurdles.

Complementing Vallée's focus on multimodal data integration [5], a recent overview by in [16] provides a detailed examination of enabling technologies, reference architectures, and application areas for healthcare DTs. The study categorizes implementations into patient monitoring, predictive diagnostics, and personalized treatment planning, while emphasizing the persistent barriers of interoperability, privacy, scalability, and rigorous validation. These observations further reinforce the necessity for semantically enriched, patient-centric DT frameworks capable of integrating multimodal data and supporting advanced analytics.

Several initiatives have explored DTs for specific organs or systems, such as the heart (Siemens Healthineers, Dassault Living Heart), brain modelling (Hewlett Packard Enterprise), and the artificial pancreas [17]. Other examples include pediatric cardiac DTs [18], a DT for the human head to detect carotid stenosis [19], a model for mitral valve disease progression [20], and condition-specific systems for ischemic stroke [20], precision cardiology [21], and AI-generated in silico intervention strategies [22]. Commercial device manufacturers are increasingly using DTs for patient-specific device design [23, 24], while imaging-based DTs support surgical planning. *However, as reported across these studies, most existing implementations remain organ-specific or condition-specific, addressing isolated physiological parameters while neglecting psychosocial, behavioral, and genomic dimensions. Such compartmentalized models cannot provide the longitudinal, whole-patient perspective required for proactive and personalized chronic care.*

As noted by Vallée [5], the ability of digital twins to integrate multimodal data streams and generate holistic patient models is essential for effective chronic care, particularly for continuous monitoring and early intervention.

Other DT-related research has addressed predictive analytics and interoperability challenges at the systems level: Jain et al. [7] developed an HPCC-based framework for COPD readmission risk prediction; Lopez-Perez et al. [8] built BD2Decide, a big-data platform for personalized head and neck cancer decision support; Satti et al. [11] created the Ubiquitous Health Profile for interoperable health data curation; and Jing et al. [10] applied fuzzy prediction for surgical planning in total knee replacement. While these works advance analytics and integration, they remain condition-specific and do not deliver unified, ontology-based patient twins.

Recent years have seen a surge in interest in applying ontology-driven and semantic approaches to DTs to achieve interoperability, enable automated reasoning, and promote reusability across heterogeneous health data sources. However, most existing ontology-based initiatives remain limited to narrow domains or specific data types, stopping short of delivering a truly holistic patient model. Xames and Topcu [25], in their comprehensive systematic literature review of DT research for healthcare systems, identified persistent gaps in scalability, interoperability, and clinical validation. Crucially, they concluded that despite promising applications, there is still no ontology framework that formally integrates diverse modalities of patient data, physiological, psychosocial, behavioral, and genomic, into a single, reusable semantic backbone suitable for advanced, privacy-preserving analytics. Balasubramanyam et al. [26] reviewed the transformative potential of DTs across healthcare domains, emphasizing predictive modelling, real-time monitoring, and patient-specific interventions. However, their survey also highlighted the fragmented nature of existing systems, the absence of standardized modelling approaches, and the need for formal semantic structures to support interoperability and reasoning. Similarly, Maleki et al. [27] introduced an intelligent DT framework integrating AI-driven analytics, IoT-based continuous monitoring, and real-time feedback loops to support adaptive and personalized care. While their architecture addresses predictive modelling and system adaptability, it lacks a formalized ontology-based backbone for semantic interoperability and reasoning, which is the core contribution of the present work. Shahzad et al. [28] proposed an ontology-driven smart health service integration framework leveraging semantic web technologies to unify disparate healthcare services. While demonstrating improved interoperability, their approach does not target the creation of a holistic, patient-level DT that spans multiple physiological, psychosocial, and genomic data sources. Another relevant contribution is presented by [29], who proposed a layered digital twin framework for healthcare that integrates IoT-based patient monitoring with electronic health records using interoperability standards such as HL7 FHIR. Their architecture enables continuous data acquisition, semantic modeling, and AI-driven predictive analytics for personalized care recommendations. While sharing conceptual similarities with the PMDT, particularly in its focus on interoperability and predictive modeling, this approach does not employ a formal OWL-based ontology nor address the integration of comprehensive psychosocial and genomic patient data as in the present work.

Thomsen et al. [30] explored ontology-driven DTs for hospital building operations, illustrating how semantic modelling can optimize recommissioning processes. Though the domain differs, their work reinforces the utility of ontologies in structuring complex, heterogeneous data for decision support, a principle directly applicable to patient-focused DTs. Riccardo [31], in a doctoral thesis, investigated ontology-based approaches in healthcare DTs, advocating for OWL-based representations as a means to formalize patient models and enable reasoning over multimodal health data. This aligns closely with the conceptual goals of the present work but stops short of delivering a complete, operationalized PMDT architecture.

*Table 1 Comparative gap analysis of prior DT approaches in healthcare, mapping categories of existing work to their limitations and showing how the Patient Medical Digital Twin (PMDT) addresses these gaps.*

| Category of Prior Work | Representative Studies | Limitations in Prior Work | How PMDT Addresses Them |
|---|---|---|---|
| **Organ-Specific Digital Twins** | Siemens Healthineers, Dassault Living Heart, [15-20] | Narrow scope focused on single organs or diseases; limited integration of psychosocial, behavioral, and genomic data; lack of whole-patient perspective. | Provides a unified, whole-patient model that integrates physiological, psychosocial, behavioral, and genomic data; supports longitudinal care and proactive interventions. |
| **Predictive Analytics Frameworks** | Jain et al. [5], Lopez-Perez et al. [6], Satti et al. [9], Jing et al. [8] | Condition-specific analytics; limited interoperability; absence of formal ontology backbone; no unified multimodal patient model. | Combines predictive analytics capability with a reusable semantic backbone in OWL; enables interoperability across heterogeneous sources; supports privacy-preserving analytics. |
| **Ontology-Driven Systems** | Xames & Topcu [23], Balasubramanyam et al. [24], Maleki et al. [25], Shahzad et al. [26], Thomsen et al. [28], Riccardo [29] | Often domain- or dataset-specific; lack of fully operational OWL-based models; limited multimodality; no real-world validated, patient-centric DTs. | Delivers a comprehensive OWL-based patient DT integrating multimodal data; validated in real-world clinical settings; extensible for diverse analytics use cases. |

**Gap**: Across these studies, no comprehensive patient DT framework exists that (i) is formally modelled as an ontology in OWL, (ii) integrates multimodal health data, spanning physiological, psychosocial, and genomic dimensions, and (iii) serves as an extensible semantic backbone for diverse analytics use cases. Most prior works remain limited to specific health conditions, isolated organ models, or narrow interoperability solutions, without addressing the need for a unified, patient-centric approach that enables both proactive care and privacy-preserving analytics across institutions. Even the most recent syntheses acknowledge that, despite advances in semantic modelling, no existing solution has demonstrated a fully operational, OWL-based patient digital twin rigorously evaluated through expert validation and structured feedback, precisely the focus of the PMDT presented in this work.

To further clarify how the proposed PMDT addresses the shortcomings of existing approaches, Table 1 provides a comparative gap analysis. It maps three major categories of prior work, organ-specific DTs, predictive analytics frameworks, and ontology-driven systems, to their key limitations, and contrasts them with the capabilities introduced by PMDT. This structured overview reinforces the unique positioning of PMDT as a holistic, OWL-based, multimodal, and privacy-preserving patient digital twin framework.

**Contribution of this paper**: This paper addresses the identified gap by introducing the Patient Medical Digital Twin (PMDT), a formally modelled, extensible, and iteratively validated in silico patient ontology, implemented in OWL. The PMDT integrates multimodal patient data, including, clinical, psychosocial, behavioral, and genomic, within a semantically enriched framework that supports automated reasoning, interoperability across heterogeneous systems, and reusability in diverse clinical contexts. Its conceptualization and modeling were refined through iterative validation with medical experts in the EU-funded QUALITOP project, ensuring clinical relevance and feasibility. While the federated architecture (Section 5) demonstrates how the PMDT can underpin advanced analytics, including predictive modelling and privacy-preserving data sharing, this paper focuses on the *conceptualization, ontology design, and validation of the semantic framework*. Broader technical implementation details and large-scale analytics evaluation are reserved for future studies.

# 3 The PMDT Vision

The Patient Medical Digital Twin (PMDT) is conceived as a next-generation, ontology-driven in silico patient model that unifies heterogeneous health data into a semantically enriched, extensible, and privacy-preserving ecosystem. Building upon the gaps identified in Section 2, the PMDT addresses the fragmentation of current healthcare digital twin efforts by integrating physiological, psychosocial, behavioral, and genomic data streams into a cohesive longitudinal patient representation.

## 3.1 Concept Overview

The PMDT is designed as a living, adaptive representation of a patient, evolving in near real time with data from electronic health records (EHRs), wearable sensors, patient-generated health data (PGHD), imaging systems, and genomic sequencing. Unlike organ- or disease-specific twins, the PMDT offers a whole-patient perspective, enabling both precision interventions and preventive strategies. This vision rests on three core design principles:

- **Formal Ontology Modeling** in the Web Ontology Language (OWL) to ensure semantic interoperability, automated reasoning, and extensibility.
- **Federated Analytics Readiness** to enable multi-institutional collaboration without the need to centralize sensitive patient data.
- **Holistic Multimodality** through the integration of biomedical, psychosocial, behavioral, and genetic determinants of health.

The conceptual architecture of the PMDT (Fig. 1) embodies a *multi-layered framework* that operationalizes these principles:

1. **Stakeholder View Level**: Role-specific interfaces for clinicians, patients, dieticians, and social workers, providing tailored visualizations of PMDT outputs.
2. **Digital Twin, AI & Big Data Technologies Level**: "Blueprints" for patient, disease, pathway, and treatment performance, enriched with risk models for diagnostics, prognostics, and decision support.
3. **Transformation & Orchestration Level**: AI services and big data pipelines for integrating and processing diverse datasets.

4. **Data Source Level**: Federated metadata repositories (EHR, PGHD, contextual embeddings) connected via secure, standards-based protocols (e.g., SMART on FHIR).

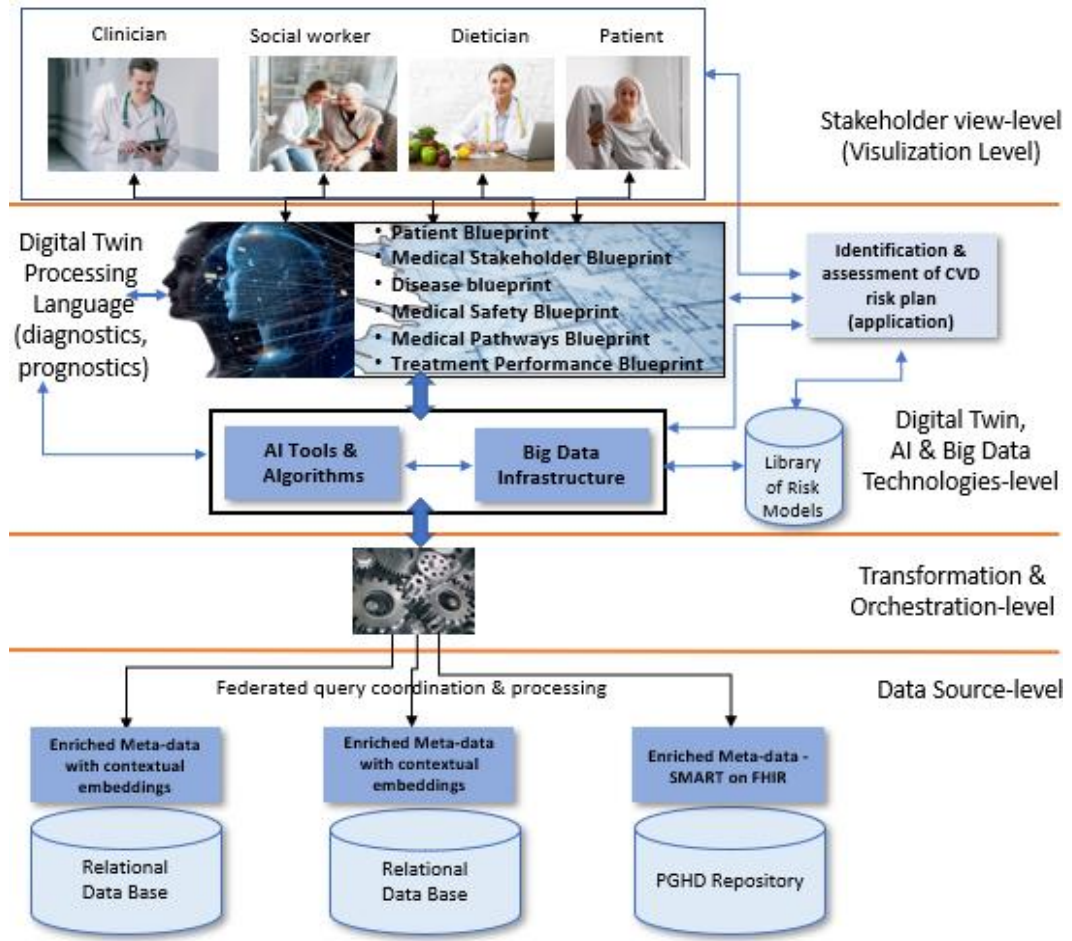

*Fig. 1 Vision for the Patient Medical Digital Twin (PMDT), showing its multi-layered architecture from data sources to stakeholder views.*

This layered structure ensures that insights from predictive analytics and ontology-based reasoning are continuously fed back to stakeholders, fostering proactive, evidence-based chronic care management.

### 3.2 Requirements for Realizing the PMDT

From the literature synthesis and stakeholder consultations, the following high-level requirements emerge:

1. **Semantic Interoperability**: adoption of standardized vocabularies and ontological structures for unambiguous data exchange.
2. **Multimodal Data Integration**: harmonization of structured, semi-structured, and unstructured health data into a unified patient model.

3. **Dynamic Synchronization**: continuous model updates in response to new clinical, behavioral, and environmental data.
4. **Predictive & Prescriptive Analytics Support**: embedding reasoning mechanisms to forecast disease trajectories and suggest interventions.
5. **Privacy-Preserving Architecture**: support for federated learning and analytics in compliance with GDPR/HIPAA.
6. **Extensibility & Reusability**: a modular structure allowing adaptation to new diseases, data types, and analytics services.

Together, these requirements directly address the specific shortcomings identified in Section 2, namely: the lack of a holistic, ontology-based patient twin that unifies multimodal health data, supports scalable interoperability, and enables privacy-preserving, clinically validated analytics across institutions.

### 3.3 Illustrative Personas and Scenarios

To ground the PMDT vision in realistic clinical contexts, we define representative personas centered on **melanoma patients**, reflecting varying disease stages and comorbidities. These scenarios will be referenced throughout the paper to illustrate how the PMDT addresses requirements outlined in Section 3.2.

**Persona 1 – "Elena" (Early-Stage Melanoma with Diabetes)**

Elena, a 54-year-old woman, has been diagnosed with early-stage cutaneous melanoma and also lives with type 2 diabetes. Her care pathway includes *surgery* and adjuvant *medication*, followed by structured *follow-ups* (surgical wound checks, glucose monitoring, and medication adherence tracking).

*Linked requirements*: *Semantic Interoperability* (harmonizing EHR records with glucose sensor data); *Multimodal Data Integration* (combining clinical, metabolic, and behavioral information); and *Predictive & Prescriptive Analytics Support* (forecasting recurrence and guiding follow-up intensity).

**Persona 2 – "Markus" (Advanced Melanoma under Immunotherapy)**

Markus, a 62-year-old man, is undergoing immunotherapy for metastatic melanoma while also experiencing hypertension. His treatment journey is characterized by evolving *disease states*, *treatment response*, and *quality of life* variations over time.

*Linked requirements*: *Dynamic Synchronization* (continuous updates with therapy cycles and imaging result); *Privacy-Preserving Architecture* (enabling multi-institutional sharing of Markus' immunotherapy records); and *Predictive &*

*Prescriptive Analytics Support* (simulating likely adverse events and disease progression).

**Persona 3 – "Aisha" (Genomic Risk for Melanoma, Family History)**

Aisha, a 39-year-old woman, carries a germline mutation associated with melanoma susceptibility and a family history of the disease. Although asymptomatic, her PMDT aggregates heterogeneous *PatientData*, including genomic risk markers, family history records, skin imaging, lifestyle indicators (e.g., sun exposure, smoking, and diet adherence), and psychosocial assessments (stress levels, social support). These data are contextualized through her *Diagnosis* of genetic susceptibility and linked to relevant *MedicalStakeholders*, including her dermatologist, genetic counselor, and primary care physician.

*Linked requirements*: *Semantic Interoperability* (integrating genomic, imaging, lifestyle, and psychosocial data within a unified ontology); *Multimodal Data Integration* (capturing behavioral, psychosocial, and clinical determinants alongside genomic risk); *Dynamic Synchronization* (updating Aisha's preventive trajectory as new lifestyle, imaging, or psychosocial data become available); *Extensibility & Reusability* (adapting the ontology for preventive oncology use cases and comorbidity-aware risk monitoring).

Together, these personas demonstrate how the PMDT provides a whole-patient perspective for melanoma care, spanning prevention, treatment, and survivorship, while addressing the technical requirements necessary for a robust, ontology-driven digital twin ecosystem.

**3.4 Positioning the PMDT in the Healthcare Ecosystem**

The PMDT is not intended as a standalone research artifact but as an integral component within a broader healthcare information infrastructure. It is designed to interface seamlessly with existing hospital information systems, national health registries, and patient-facing applications, while also enabling cross-institutional collaboration through privacy-preserving data sharing mechanisms.

At the micro level, the PMDT provides *personalized decision support*, delivering predictive insights, simulation-based treatment planning, and continuous condition monitoring for individual patients. At the meso level, it supports *care coordination* across multidisciplinary teams by offering role-specific visualizations and shared context. At the macro level, federated deployments of PMDT instances enable *population health analytics* without compromising patient privacy, fostering the creation of learning healthcare systems.

The architecture's semantic backbone ensures that data and models remain *extensible*, supporting emerging diseases, new data modalities, and evolving regulatory requirements. This positions the PMDT not only as a clinical decision-support tool but as a *strategic enabler* of digital transformation in chronic care management.

# 4 PMDT Design and Formal Modeling

The PMDT translates the conceptual vision outlined in Section 3 into a formal, operational design that emphasizes semantic rigor, modularity, and extensibility. Unlike the high-level architectural layers previously discussed, this section focuses on the internal design mechanisms, specifically the blueprint structures, ontology modeling approach, and knowledge representation techniques, that instantiate the PMDT.

By grounding the framework in the Web Ontology Language (OWL), the design ensures precise semantics, automated reasoning, and reusability across diverse healthcare contexts. The blueprint structures formalize core categories of multimodal health data (e.g., patient, disease, treatment, pathways, performance, safety, stakeholders) into modular, semantically linked knowledge components. These blueprints provide the foundation for interoperable and extensible patient representations, enabling both individual-level analytics and federated, privacy-preserving collaboration.

The design process followed an iterative, expert-driven methodology. Prototypes and ontology drafts were repeatedly refined in validation workshops with clinicians, data scientists, and system engineers from partner medical institutions. This ensured that each modeling decision was aligned with both the research gaps identified in Section 2 and the functional requirements outlined in Section 3.2.

## 4.1 Overall Architecture

As introduced in Section 3, the PMDT follows a layered architecture, connecting data sources, orchestration pipelines, knowledge structures, and stakeholder interfaces. Within this stack, the Digital Twin & Knowledge Layer plays a central role: it structures heterogeneous health information into semantically coherent, reusable, and extensible components referred to as *Blueprints*.

The PMDT Blueprints (Fig. 2) define seven interconnected knowledge types, each capturing a distinct aspect of the patient model:

1. **Medical Stakeholder Blueprint**: represents the capabilities of health professionals and institutions involved in patient care, including clinicians, nurses, social workers, and nutritionists, along with their skills, responsibilities, and available facilities.
2. **Patient Blueprint**: encapsulates the longitudinal patient profile, covering demographics, treatment history, lifestyle, psychosocial factors, and patient-generated data.
3. **Disease Blueprint**: describes disease-specific information such as diagnosis, biomarkers, pathology, symptoms, comorbidities, and quality-of-life impact. For oncology, this includes cancer subtypes, tumor markers, and genetic test results.
4. **Treatment Blueprint**: captures current and historical treatment interventions, including therapies, medications, and surgical procedures.
5. **Treatment Performance Blueprint**: monitors performance and outcome indicators, such as readmission rates, workforce utilization, timeliness of care, patient experience, and overall cost-effectiveness.
6. **Medical Safety Blueprint**: encodes clinical practice guidelines, safety procedures, and data governance rules (privacy, security, regulatory compliance).
7. **Medical Pathways Blueprint**: defines treatment plans, schedules, and care processes. While this blueprint can be extended to support executable workflow representations using standard notations such as BPMN 2.0, this work is currently ongoing and not yet validated. Nevertheless, it highlights the extensibility of the PMDT in capturing both declarative knowledge (ontologies) and procedural knowledge (care pathways).

Together, these Blueprints ensure that patient, disease, treatment, and context-specific knowledge is captured in a modular yet interconnected way, enabling semantic interoperability, automated reasoning, and reuse across clinical contexts. They directly address the requirements outlined in Section 3.2: interoperability, multimodal integration, dynamic synchronization, privacy-preservation, and extensibility.

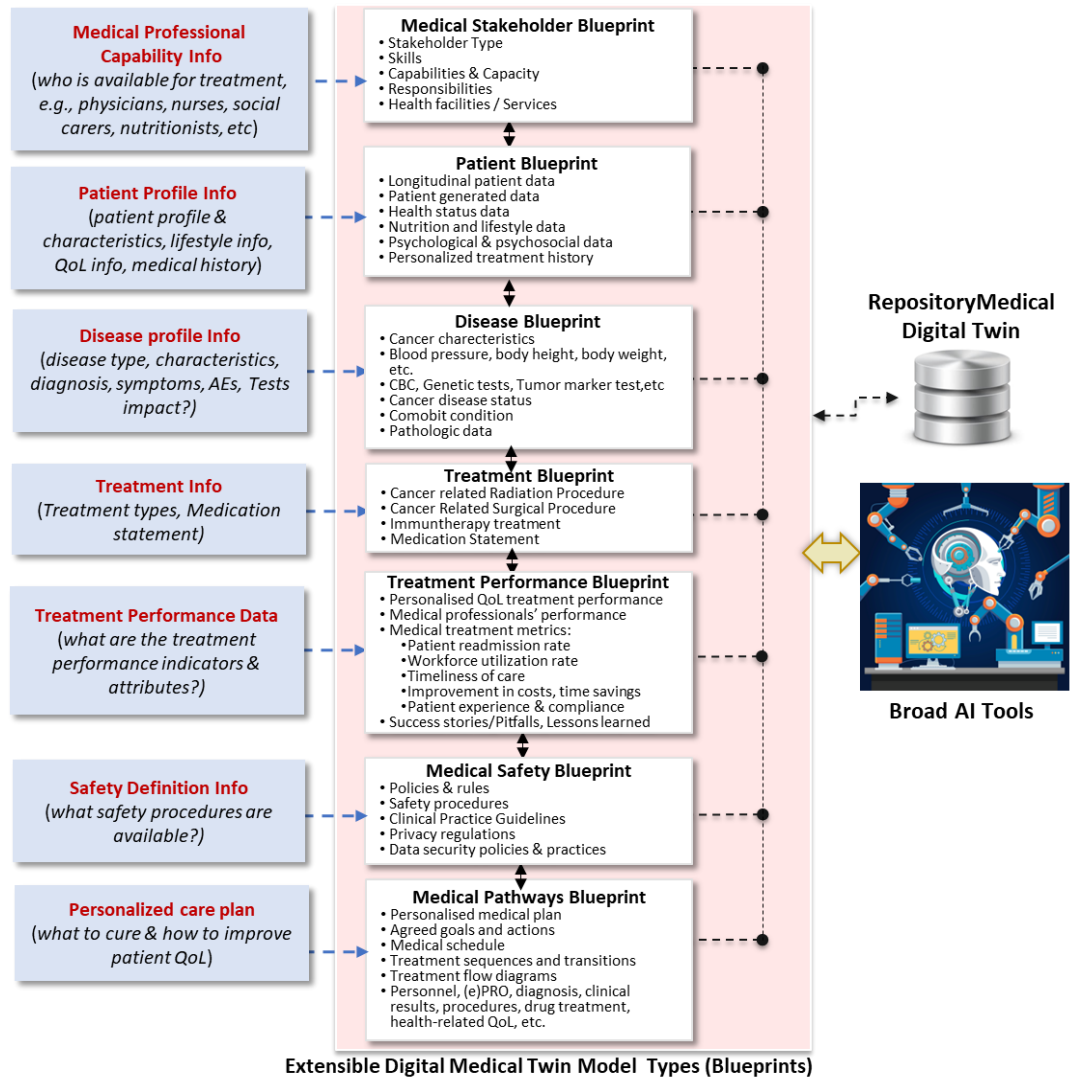

*Fig. 2 Conceptual blueprint structures of the Patient Medical Digital Twin (PMDT), representing seven abstract knowledge types that capture core categories of multimodal health data.*

In the next subsection (Section 4.2), we demonstrate how these abstract Blueprints are instantiated into a formal ontology model in OWL, with detailed UML representations of their classes, relations, and constraints.

## 4.2 Ontology model in OWL

The PMDT ontology provides the formal OWL specification of the Blueprint structures introduced in Section 4.1. While the Blueprints capture the conceptual foundations of the patient model, their instantiation in OWL enables precise semantics, automated reasoning, and interoperability across heterogeneous data sources. To ensure both modularity and expressiveness, the ontology is organized into a core *Main View*, which integrates patients, diseases, treatments, data, and stakeholders, and several specialized views that elaborate on critical dimensions such as *Trajectory, Treatment and Follow-Up, Medical Safety, Medical Pathways, and Adverse Events*. This section presents the methodology used to develop the ontology (Section 4.2.1) and the conceptual views that structure its design (Section 4.2.2), each illustrated with personas and competency questions.

### 4.2.1 Ontology Development Methodology

The PMDT ontology was developed using the knowledge-engineering methodology in [32], following iterative design, evaluation, and refinement cycles. The process included (i) scoping through competency questions, (ii) iterative modeling using OWL, and (iii) continuous validation through feedback workshops with clinicians, data scientists, and medical partners in the QUALITOP project. This ensured alignment with the requirements defined in Section 3.2 and the representative scenarios of Section 3.3.

Competency questions were central to guiding the ontology's scope and coverage. At a high level, they focused on:

- **Personalized care planning**, e.g., "What care plan should be recommended for a melanoma patient with comorbid cardiovascular disease?"
- **Treatment decision support**, e.g., "Which treatment options exist for a patient with Grade 2 cytokine release syndrome after CAR-T infusion?"
- **Multimodal determinants of outcome**, e.g., "How does a patient's lifestyle (smoking, alcohol) affect disease progression and treatment response?"
- **Longitudinal quality-of-life assessment**, e.g., "What is the expected quality of life of a melanoma patient one year after CAR-T infusion?"

These overarching questions shaped the ontology's structure, while more detailed, view-specific competency questions are provided in Section 4.2.2 to illustrate the reasoning supported by each conceptual view. While these competency questions primarily reflect the *descriptive reasoning* enabled directly by the ontology, the PMDT framework extends further: when embedded in the implementation architecture (Section 5) and combined with AI/ML techniques, it supports *predictive and prescriptive analytics*. This enables advanced use cases such as forecasting the likelihood of immune-related adverse events, simulating long-term quality-of-life trajectories, or recommending adaptive treatment adjustments. The evaluation of these extended capabilities is presented in Section 6.

### *4.2.2 Ontology Conceptual Model*

The PMDT ontology builds on the Blueprints introduced in Section 4.1. To manage complexity, it is presented through multiple conceptual views, each highlighting a core dimension of the patient model. In OWL, each Blueprint is represented by dedicated classes, object properties, and restrictions, enabling formal reasoning and semantic querying.

**Main View**

Fig. 3 presents the *Main View* of the PMDT ontology. This view acts as the central integration layer, showing how patients, diseases, treatments, data, stakeholders, and outcomes are interlinked. From here, the ontology branches into the specialized views described in the following subsections:

1. **Treatment, Treatment Performance, and Follow-Up View**: models how interventions are carried out, evaluated, and monitored across the care pathway.
2. **Trajectory View**: captures the temporal evolution of disease, treatment, and quality-of-life states.
3. **Medical Safety View**: formalizes clinical guidelines, safety procedures, and data governance rules.
4. **Medical Pathways View**: represents structured workflows, pathway steps, and clinical goals.
5. **Adverse Events View**: details the short- and long-term toxicities and complications associated with cancer therapies.

At the center of the Main View is the *Patient* class, which provides the entry point for reasoning across all views. Patients are connected to:

- **Diagnosis**: linking both cancer-specific conditions (via the PatientCancerDisease subclass) and comorbidities represented through the Disease Ontology (DO). This provides the semantic foundation for multimorbidity representation.

- **Treatment**: connected to *TreatmentPerformance* and *Follow-Up*, forming the bridge to the next dedicated sub-section.
- **AdverseEvents**: linked to treatments, diagnoses, and follow-ups, preparing the ground for the *Adverse Events View*.
- **PatientData**: serves as the *multimodal integration hub*, consolidating heterogeneous datasets into a unified patient representation. It captures structured clinical data (e.g., medical history, test results), patient-reported outcomes (quality of life scores, psychosocial assessments), and behavioral/lifestyle factors (nutrition, physical activity, smoking, alcohol consumption, sleep). By explicitly modeling these modalities, PatientData supports longitudinal reasoning in the *Trajectory View*, risk detection and governance in the *Medical Safety View*, and personalization of *Treatment and Follow-Up* plans. This multimodal design ensures that both biomedical signals and patient-centered perspectives are equally represented in decision support.aggregating multimodal datasets (clinical history, lifestyle, nutrition, psychosocial, psychological, and QoL data) that drive the *Trajectory* and *Safety* views.
- **MedicalStakeholders**: capturing the roles of physicians, nurses, and social workers, and serving as the connector to the *Medical Pathways View*.

Together, this Main View serves as the *unifying backbone*: every specialized view is an elaboration of one or more of these central relationships. This integrative view enables PMDT to answer complex competency questions that span multimodal data, diagnostic reasoning, and stakeholder responsibilities, such as:

***Descriptive:***

- *How can multimodal patient data (clinical history, lifestyle, nutrition, psychosocial, psychological, and QoL data) be combined to provide a holistic view of disease status and progression?*
- *Which comorbid diagnoses influence the design of individualized care plans for oncology patients?*
- *How do medical stakeholders collaborate in the follow-up pathway?*
- *How can lifestyle and psychosocial data be integrated with diagnostic evidence to personalize both treatment and long-term follow-up strategies?*

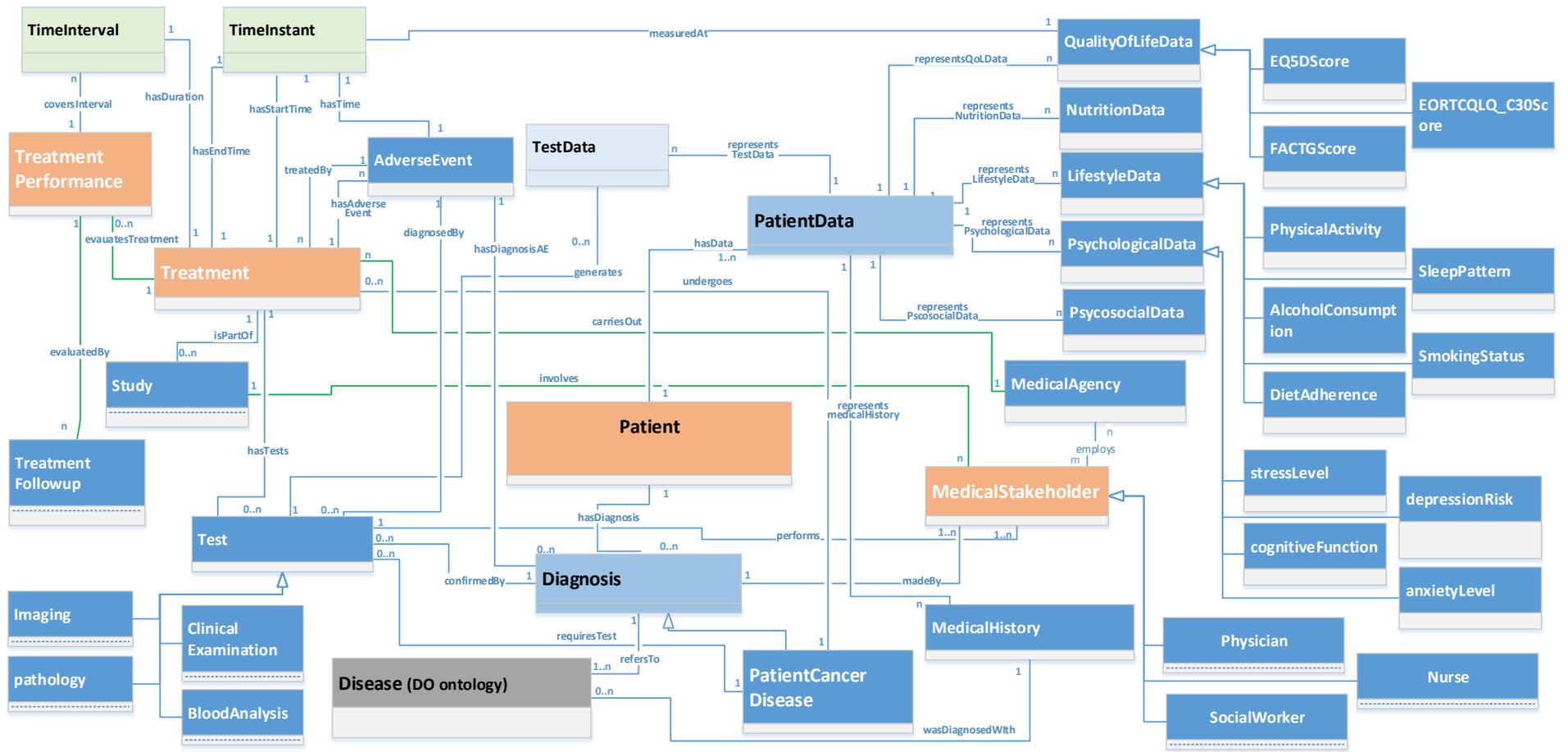

*Fig. 3 PMDT ontology conceptual model (UML class diagram), illustrating the formalization of the Blueprints in OWL.*

*Predictive/Prescriptive:*
- *How can multimodal patient data be leveraged to predict future disease trajectories or adverse event?*
- *What prescriptive recommendations (e.g., adaptive follow-up intensity, preventive interventions) can be generated to optimize long-term outcomes?*

These competency questions are directly illustrated by *Aisha* (Persona 3, Section 3.3), a 47-year-old woman with breast cancer and hypertension. Aisha's heterogeneous *PatientData* includes blood pressure monitoring and tumor imaging, lifestyle indicators such as diet adherence and smoking status, and psychosocial assessments of stress and support networks. Through the PMDT ontology, this data is linked to her *Diagnosis* and coordinated across *MedicalStakeholders* (oncologist, cardiologist, social worker). This allows queries such as: *"How should Aisha's cancer treatment plan be adapted in light of her comorbid hypertension, and which stakeholder is accountable for each dimension of her care?"*

**Treatment, Treatment Performance, and Follow-Up View**

Fig. 4 illustrates the Treatment, Treatment Performance, and Follow-up View of the PMDT ontology. This view formalizes how interventions are represented, evaluated, and continuously monitored across the patient trajectory.

The *Treatment hierarchy* distinguishes major intervention types (surgery, medication, vaccination, and cancer therapy), with the latter further refined into immunotherapy, chemotherapy, and radiation therapy. This modular structure ensures extensibility, enabling the ontology to accommodate emerging modalities such as cell-based therapies.

To capture the *evaluation of treatments*, the ontology introduces the *TreatmentPerformance* class. Each performance evaluation is associated with a specific treatment (evaluatesTreatment) and may be linked to a disease context (associatedWithDisease). Treatment performance is decomposed into three complementary metric categories (hasMetric):

- **EfficiencyMetric**: captures resource-oriented indicators, such as workforce utilization and waiting times.

- **EffectivenessMetric**: models clinical outcomes, such as survival rates, remission, or recurrence.
- **PatientCentredOutcome**: formalizes patient-reported outcomes, including satisfaction, adherence, or quality-of-life assessments (basedOnQoL).

To ensure longitudinal tracking, each performance evaluation is anchored to a *time instant* or *time interval* (measuredAt), and may be integrated into a broader *trajectory* (partOfTrajectory), allowing reasoning over temporal trends (check Trajectory view next).

The ontology also models *Treatment Follow-up*, representing post-intervention monitoring. Specialized subclasses (e.g., SurgeryFollowup, MedicationFollowup, CancerTherapyFollowup) ensure that follow-ups remain treatment-specific, reducing ambiguity between different care processes. Each follow-up is explicitly linked to the treatment it concerns (followUpOfSurgery, followUpOfMedication, etc.), and is associated with one or more performance evaluations (hasPerformanceEvaluation). Furthermore, follow-ups can directly report patient-centred outcomes (reportsOutcome), such as satisfaction surveys or self-reported side effects, thus bridging clinical and patient-generated perspectives.

Together, this view integrates *treatment actions, follow-up processes, and performance evaluations* into a coherent semantic model. It enables competency queries such as:

***Descriptive:***

- *Which efficiency and effectiveness metrics were recorded for a patient's immunotherapy, and at what time intervals?*
- *What patient-reported outcomes were collected during the 3-month follow-up after surgery?*
- *How does treatment performance evolve across repeated follow-ups along the patient's trajectory?*

***Predictive/Prescriptive:***

- *Based on past treatment performance and multimodal patient data, what is the predicted effectiveness or likelihood of adverse events for a given treatment?*
- *Which prescriptive adjustments (e.g., dose modification, earlier follow-up, switching therapy) should be recommended?*

This makes the ontology not only descriptive but also evaluative, providing a foundation for *predictive analytics and adaptive decision support* in melanoma care.

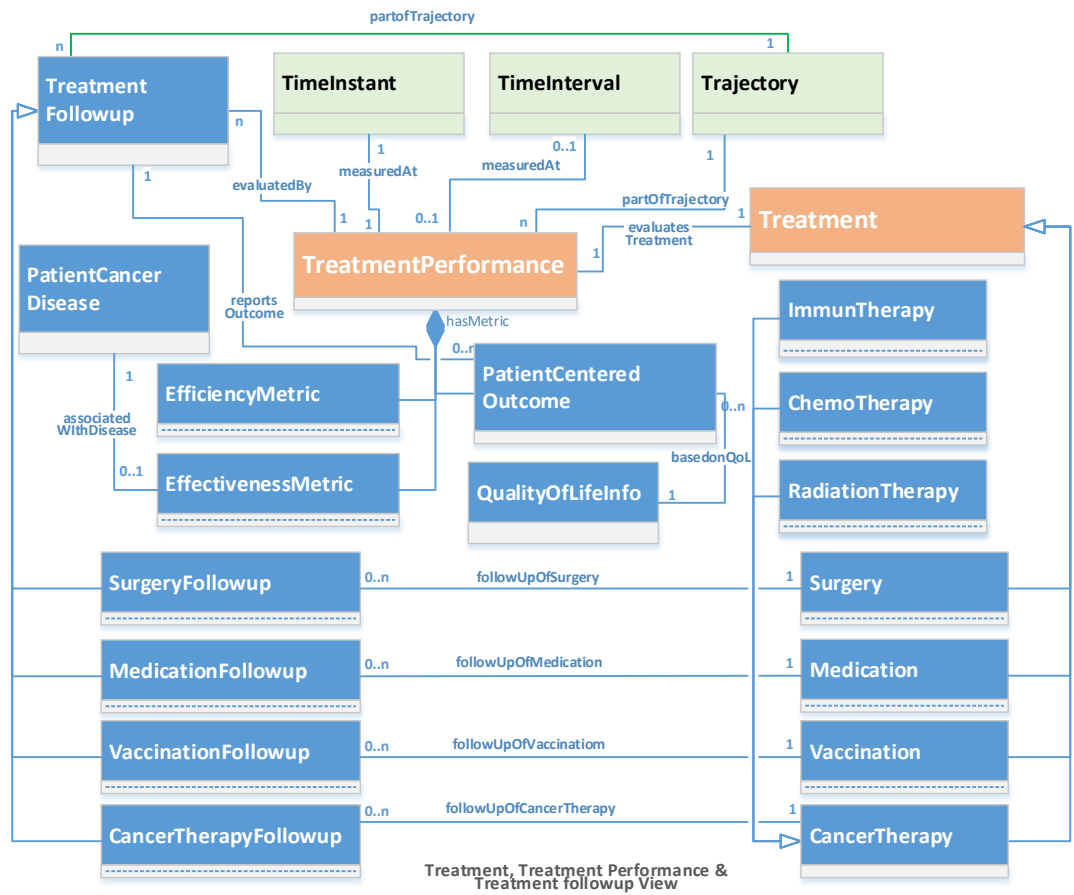

*Fig. 4 PMDT ontology – Treatment, Performance, and Follow-up View*

This view is directly illustrated by the case of *Elena* (Persona 1, Section 3.3). Elena's melanoma treatment pathway combines surgery and medication, which are modeled as subclasses of Treatment. Her subsequent follow-ups are captured through SurgeryFollowup and MedicationFollowup, with associated TreatmentPerformance indicators such as wound healing rates and glycemic stability (reflecting her comorbid diabetes). PatientCentredOutcome captures her reported satisfaction and recovery experience. By linking follow-ups to concrete outcomes, the ontology supports longitudinal evaluation of treatment effectiveness and patient well-being, addressing the requirements of Semantic Interoperability, Multimodal Data Integration, and Predictive & Prescriptive Analytics Support.

**Trajectory View**

A core requirement of the PMDT is to represent the *temporal evolution* of patient states across disease, treatment, and quality-of-life dimensions. The Trajectory View (Fig. 5) formalizes this longitudinal aspect by introducing the class *Trajectory* as a container for temporally ordered states.

Each trajectory is specialized into three main categories:

- **PatientCancerDiseaseTrajectory**: capturing disease evolution, including recurrence or remission events.
- **TreatmentTrajectory**: representing treatment phases, changes in regimen, or cycles of administration.
- **QoLTrajectory**: describing the evolution of patient-reported and clinically assessed quality-of-life measures.

Within each trajectory, instances of *State* represent snapshots in time. These states are linked to *TimeInstant* or *TimeInterval*, allowing reasoning about both point events (e.g., a single adverse event) and extended phases (e.g., chemotherapy cycle). The ontology defines subtypes of *State* to capture domain-specific aspects:

- *TreatmentState*: describes the patient's treatment status at a given interval (e.g., on immunotherapy, post-surgery recovery).
- *PatientCancerDiseaseState*: describes disease progression (e.g., stable, partial response, progression).
- *QoLMeasurementState*: captures quality-of-life assessment results, linked to *QualityOfLifeInfo*.

To represent temporal aspects of disease, treatment, and quality-of-life trajectories, PMDT reuses the *OWL-Time ontology* (W3C Recommendation [4]). The ontology defines standardized constructs such as time:Instant and time:Interval, which we align to our classes TimeInstant and TimeInterval. This ensures interoperability with external biomedical and clinical datasets that adopt OWL-Time for temporal reasoning.

Transitions between states are ordered via the properties *precedes* and *follows*, enabling representation of *trajectories over time*. This supports competency questions such as:

*Descriptive:*

- *Compare a patient's QoL before and 3 months after CAR-T infusion.*

- *Trace the disease trajectory from diagnosis through relapse.*

**Predictive & Prescriptive:**
- *Given multimodal patient data and past trajectories, what is the likely disease progression or remission pattern for a new patient?*
- *Based on predicted trajectories, which intervention strategy (e.g., intensified monitoring, change of therapy, supportive care) would optimize long-term outcomes?*

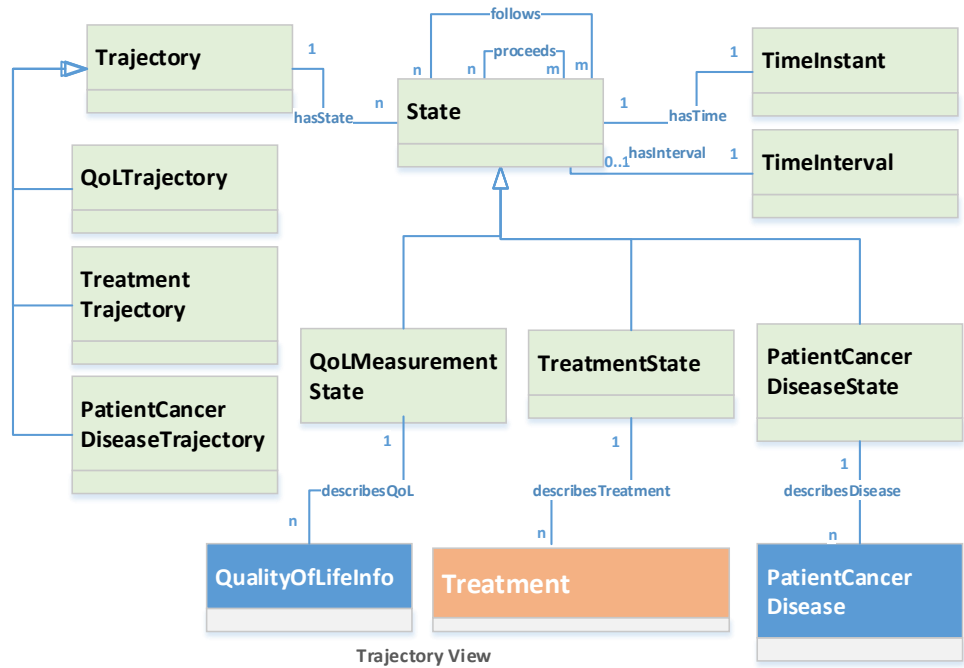

*Fig. 5 PMDT ontology – Trajectory View*

This view is exemplified by *Markus* (Persona 2, Section 3.3), a 62-year-old patient undergoing immunotherapy for advanced melanoma. Markus's *TreatmentTrajectory* captures successive cycles of immunotherapy, represented as ordered *TreatmentStates*. Parallel to this, his *PatientCancerDiseaseTrajectory* records partial remission followed by immune-related complications. His *QoLTrajectory* links quality-of-life measurements across time, reflecting fatigue and hypertension. By explicitly modeling trajectories, the PMDT can represent Markus's clinical journey as a temporally ordered set of states, fulfilling requirements of Dynamic Synchronization, Predictive Analytics, and Multimodal Data Integration.

**Medical Safety View**

Fig. 6 illustrates the Medical Safety view of the PMDT ontology. This view formalizes the representation of rules, events, and governance mechanisms that ensure safe and compliant patient care. At its core is the *MedicalSafetyRule* class, which generalizes four key categories: *ClinicalGuideline, SafetyProcedure, PrivacyRegulation, DataSecurityPolicy*, and *AccessPolicy*. These rules regulate treatments, mitigate adverse events, and ensure compliance with medical and legal standards.

Safety events are explicitly modeled through the *SafetyEvent* class, which may involve patients and be triggered by treatments or clinical workflows. Safety events may *violate* a MedicalSafetyRule, or conversely, be *mitigated by* specific rules and procedures. This enables the ontology to capture both violations (e.g., non-compliance with infection control procedures) and safeguards (e.g., adverse event monitoring protocols).

To incorporate *data governance*, the ontology introduces the *PatientData* class. Each patient instance is linked to one or more PatientData instances, which represent their medical records, sensor data, or patient-reported outcomes. PatientData is *restrictedBy* privacy regulations (e.g., GDPR, HIPAA) and *securedBy* data security policies (e.g., encryption, access control). Access to PatientData by a MedicalStakeholder is explicitly modeled, requiring *consent* from the patient and *enforcement* of corresponding MedicalSafetyRules.

To strengthen data governance and privacy modeling, we propose an extension to the Medical Safety View with three new classes: *DataSource*, *ConsentStatement*, and *AccessPolicy*. Each *PatientData* instance may be semantically linked to its origin via hasSource (e.g., EHR, wearable, PGHD), governed by an *AccessPolicy*, and backed by an explicit *ConsentStatement* from the Patient. Notably, *AccessPolicy* is modeled as a subclass of *MedicalSafetyRule*, reflecting its regulatory role alongside *PrivacyRegulation*, *DataSecurityPolicy*, and *SafetyProcedure*. While these classes were not included in the validated QUALITOP pilot implementation, they represent a natural extension of the PMDT ontology to support reasoning over provenance, consent compliance, and rule enforcement in privacy-preserving federated analytics — aligning with GDPR and HIPAA principles..

Through these constructs, the Medical Safety view covers both *clinical safety* (guidelines, procedures, adverse event management) and *data safety* (privacy, security, governance). This enables the PMDT ontology to reason not only about what treatments are safe, but also how patient data must be protected and accessed. Importantly, PMDT distinguishes between *object-level events* (e.g., treatments, adverse events, patient data) and *meta-level rules* (e.g., safety procedures, privacy regulations) that govern their use. This separation follows the meta-logical distinction described by Hirankitti & Mai in [33], ensuring that the ontology can reason not only about clinical facts but also about the rules and constraints that regulate safe and compliant care.

This view can also be exemplified by *Markus* (Persona 2, Section 3.3), who is undergoing immunotherapy for metastatic melanoma while managing hypertension, *Medical Safety Rules* play a central role in governing his treatment journey. His PMDT integrates *SafetyProcedures* and *ClinicalGuidelines* to monitor and mitigate immune-related adverse events, while ensuring adherence to *PrivacyRegulations* and *DataSecurityPolicies* that govern the secure sharing of his multi-institutional records.

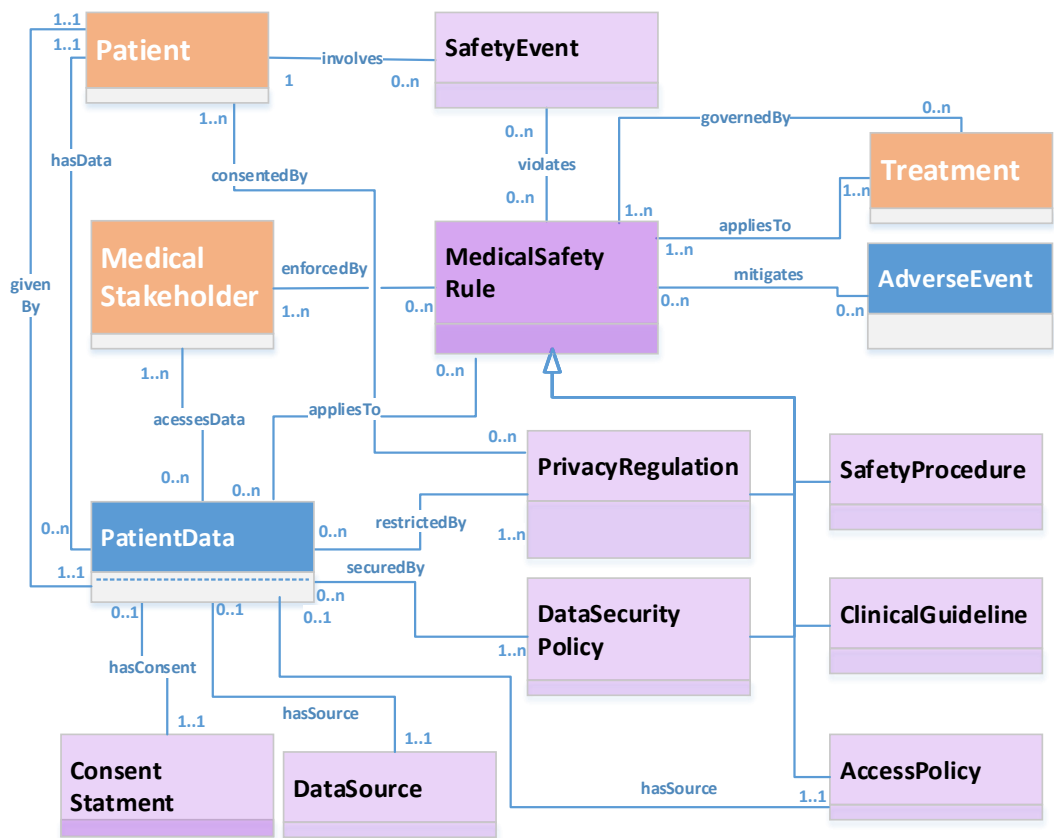

*Fig. 6 PMDT ontology – Medical Safety View*

By linking Markus' *PatientData* (including imaging, blood markers, and therapy cycles) with safety rules, the PMDT supports proactive detection of risks, enforcement of best practices, and GDPR-compliant data governance. This supports *Semantic Interoperability* (by embedding safety rules and privacy policies into the ontology for automated governance). It enables competency queries such as:

***Descriptive:***

- *Which safety procedures apply to Markus' ongoing immunotherapy treatment?*
- *Was a specific adverse event (e.g., cytokine release syndrome) associated with a violation of any clinical guideline?*
- *Which privacy regulations restrict access to Markus' genomic data shared across institutions?*
- *Which data security policies (e.g., encryption, role-based access) are required before his imaging data can be accessed by a clinical researcher?*
- *Which safety rules mitigate the risk of immune-related adverse events in melanoma patients undergoing CAR-T or immunotherapy?*

***Predictive & Prescriptive:***

- *Based on patient characteristics and historical data, what is the probability of a safety violation or adverse event occurring under a specific treatment protocol?*
- *Which preventive actions, adaptive safety rules, or governance mechanisms should be enacted to minimize risk while ensuring compliance with both clinical guidelines and data protection laws?*

**Medical Pathways View**

Fig. 7 illustrates the Medical Pathways view of the PMDT ontology. This view formalizes how structured clinical workflows and treatment plans are represented. At its core is the *MedicalPathway* class, which aggregates ordered *PathwaySteps*. Each PathwayStep may have a corresponding *TreatmentAction*, is sequenced through follows / precedes relations, and is executed by one or more *MedicalStakeholders* (e.g., surgeon, nurse, social worker).

Pathways are designed to achieve specific *ClinicalGoals*, such as tumor resection or comorbidity management, and are grounded in *ClinicalGuidelines* that encode best practices. To ensure safety and accountability, steps can be monitored by *SafetyProcedures* and their outcomes captured through links to *TreatmentPerformance* metrics. This integration ensures that pathways are not only prescriptive (defining the plan) but also evaluative (tracking effectiveness and adherence).

This view directly supports competency questions such as:

***Descriptive:***

- *Which stakeholder is responsible for executing each step of the pathway?*
- *What clinical guidelines underpin a patient's care pathway?*
- *What performance outcomes have been observed after a given pathway step?*

***Predictive & Prescriptive:***

- *How is the pathway likely to evolve for patients with similar multimodal profiles (e.g., comorbidities, lifestyle, psychosocial context)?*
- *How should the pathway be adapted in real time (e.g., changing sequence or adding additional monitoring) to optimize safety, guideline adherence, and patient-centred outcomes?*

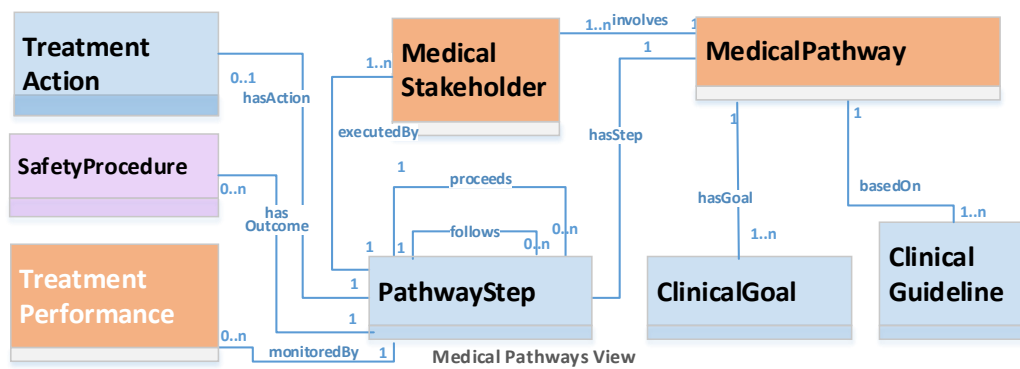

*Fig. 7 PMDT ontology – Medical Pathway View*

This view can be exemplified by *Elena (Persona 1, Section 3.3)*, a 54-year-old woman with early-stage melanoma and type 2 diabetes. Her PMDT integrates a structured pathway involving surgery, wound follow-up checks, glucose monitoring, and adherence to adjuvant medication. Each step is linked to specific stakeholders (e.g., surgeon, endocrinologist, diabetes nurse), aligned with clinical guidelines, and evaluated against treatment performance and patient-centered outcomes. By explicitly representing the sequencing, responsibilities, and goals of Elena's pathway, the PMDT enables dynamic reasoning over multimodal care plans.

**Adverse Events View**

The *Adverse Events View* ( Fig. 8) specializes the representation of *AdverseEvent* to focus on *immune-related adverse events (irAEs)*, in line with the pilot study described in Section 6.1 of the QUALITOP H2020 project that partially funds this research.

Each *AdverseEvent* is temporally anchored (*recordedAt TimeInstant/TimeInterval*), graded using CTCAE-based [34] *SeverityGrade*, and classified into irAE categories through *AdverseEventCategory*. Two major categories are distinguished:

- **Organ-specific irAEs**, such as pneumonitis, colitis, endocrinopathy, hepatitis, and myocarditis.
- **Systemic syndromes**, such as cytokine release syndrome (CRS) and immune effector cell–associated neurotoxicity syndrome (ICANS).

Causality is explicitly modeled through links to *Treatment* (*causedBy*) and *PatientCancerDisease* (*associatedWith*), while confirmation may derive from diagnostic *Tests* (e.g., imaging, blood analysis). Events may be reported either by the *Patient* or by a *MedicalStakeholder*, and then diagnosed, graded, and managed clinically. Management may involve both *MedicalStakeholders* (e.g., oncologists, nurses) and additional *Treatments* (e.g., corticosteroids for pneumonitis, hormone replacement for endocrinopathy).

To ensure safe and compliant care, each irAE is also governed by *MedicalSafetyRules* (see Medical Safety View). These rules may be violated (e.g., if recommended monitoring is missed) or mitigated (e.g., by applying established clinical guidelines for early steroid administration).

This refined model allows PMDT to support competency questions such as:

***Descriptive:***

- *Which immune-related adverse events were reported following CAR-T or checkpoint inhibitor therapy?*

- *How were irAEs managed across follow-up visits, and by which stakeholders or treatments?*
- *Which tests confirmed the diagnosis of a suspected irAE?*
- *Which safety guidelines were triggered or violated during irAE management?*

**Predictive** & **Prescriptive**:

- *Based on a patient's multimodal profile (e.g., treatment type, biomarkers, comorbidities), what is the likelihood of developing a specific irAE such as colitis or cytokine release syndrome?*
- *What preventive monitoring strategies or treatment adjustments should be recommended to minimize the risk or severity of irAEs in similar patients?*

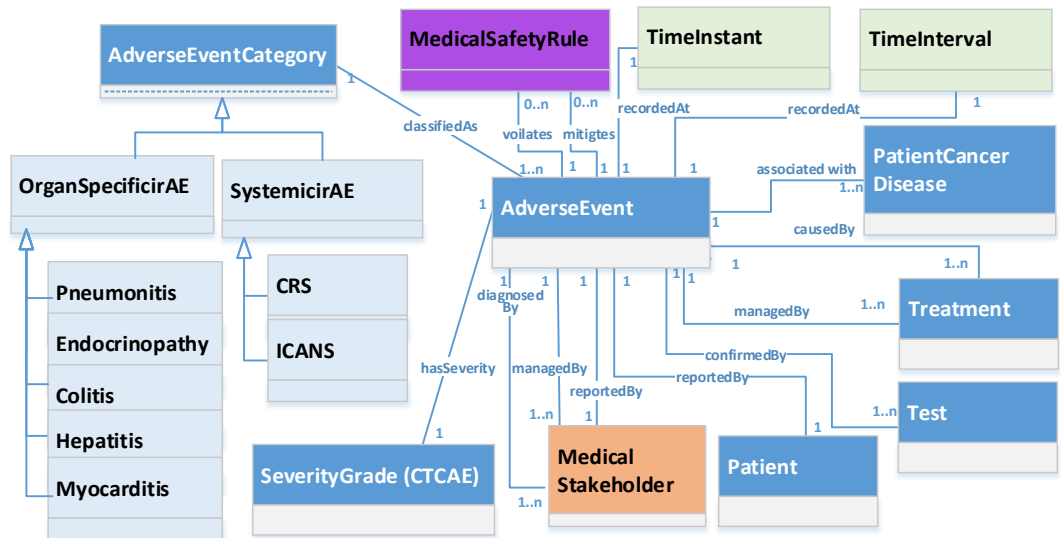

*Fig. 8 PMDT ontology – Adverse Events View*

This view can be exemplified by *Markus (Persona 2, Section 3.3)*, a 62-year-old melanoma patient undergoing immunotherapy, develops *grade 2 colitis* three months into treatment. In the PMDT ontology, this event is classified as an organ-specific irAE (*colitis*), temporally anchored at the relevant *TimeInstant*, and graded with *SeverityGrade*. The event is reported by Markus during a follow-up visit, confirmed by *Test* (blood analysis and colonoscopy), and managed by his treating oncologist (*MedicalStakeholder*) through corticosteroid administration (*Treatment*). The management process is further linked to *MedicalSafetyRules* that govern immunotherapy protocols. This representation enables longitudinal reasoning across Markus's trajectory, connecting his diagnosis, treatment, adverse events, and safety governance in a semantically integrated manner.

## 4.3 Knowledge representation

The PMDT ontology is formally implemented in the *Web Ontology Language* (OWL 2.0), enabling automated reasoning, semantic interoperability, and extensibility across heterogeneous healthcare data sources. Core modeling constructs include *classes, object properties, data properties, and restrictions*, with cardinalities specified to ensure clinically valid associations (e.g., each *Diagnosis* must concern at least one *Disease* and exactly one *Patient*). This formalization allows the ontology to support both descriptive queries and advanced inferencing, such as automatically classifying adverse events by severity or inferring follow-up requirements from treatment types.

To ensure interoperability with established biomedical standards, PMDT *reuses and aligns external ontologies where appropriate*. Cancer diseases and comorbidities are linked to the *Disease Ontology (DO)* [3], adverse events and severity grading to the *CTCAE standard* [34] (Common Terminology Criteria for Adverse Events), and temporal constructs such as *TimeInstant* and *TimeInterval* to the *W3C OWL-Time Ontology* [4]. This alignment avoids redundancy, facilitates integration with existing clinical datasets, and ensures that PMDT remains compatible with established data-sharing initiatives.

To further support semantic interoperability and integration with healthcare information systems, the ontology has been conceptually aligned with HL7 FHIR and OMOP CDM. Core PMDT classes such as *Patient*, *Diagnosis*, and *Treatment*, and *TestData* correspond to HL7 FHIR resources (e.g., Patient, Condition, Procedure, Medication), while treatment and measurement-related concepts map to OMOP CDM constructs like DrugExposure and Measurement. Although the FHIR and OMOP alignments have not yet been formally implemented in OWL, this conceptual interoperability enables future integration with clinical information systems and standardized data exchange protocols..

Patient-centered dimensions such as quality-of-life, psychosocial, and lifestyle data are represented as *modular extensions*, linked through the central *PatientData* class. This modular design supports *multimodality* by enabling the ontology to integrate clinical, genomic, behavioral, and patient-reported information under a single formal framework. By grounding the PMDT ontology in OWL 2.0 and established domain ontologies, the model achieves a balance between semantic rigor, clinical expressiveness, and extensibility toward emerging data modalities.

The conceptual model is presented through multiple views in Section 4.2.2, each grounded in these representation principles.

# 5 Implementation

The implementation of the Patient Medical Digital Twin (PMDT) ontology translates the conceptual model described in Section 4 and the multi-layered vision outlined in Section 3 into a working digital health ecosystem. It operationalizes the PMDT's core design principles, semantic interoperability, federated analytics readiness, and holistic multimodality, through a concrete software stack. The focus of the implementation is twofold: first, to realize the ontology in OWL 2.0 as a semantic backbone that ensures interoperability and reasoning across heterogeneous multimodal data sources; and second, to embed the ontology within a modular, service-oriented architecture that supports privacy-preserving federated queries and user-facing analytics. The following subsections present (i) the overall system architecture and its main components (Section 5.1), and (ii) the ontology implementation in OWL (Section 5.2). Validation and evaluation of these implementation choices, including expert feedback and analytical query execution, are discussed in Section 6.

## 5.1 Implementation Architecture

The implementation[1] of the Patient Medical Digital Twin (PMDT) ontology has been realized within the QUALITOP digital health ecosystem as a modular, service-oriented architecture (Fig. 9).

The architecture follows a layered design to ensure semantic interoperability, data privacy, and extensibility across distributed hospital environments. Four main components can be distinguished:

---

[1] Video demos of the platform are available at:
https://drive.google.com/drive/folders/1BmFfh2C1yEB5PJeCX0sdgxDdpydsEgKg?usp=sharing)

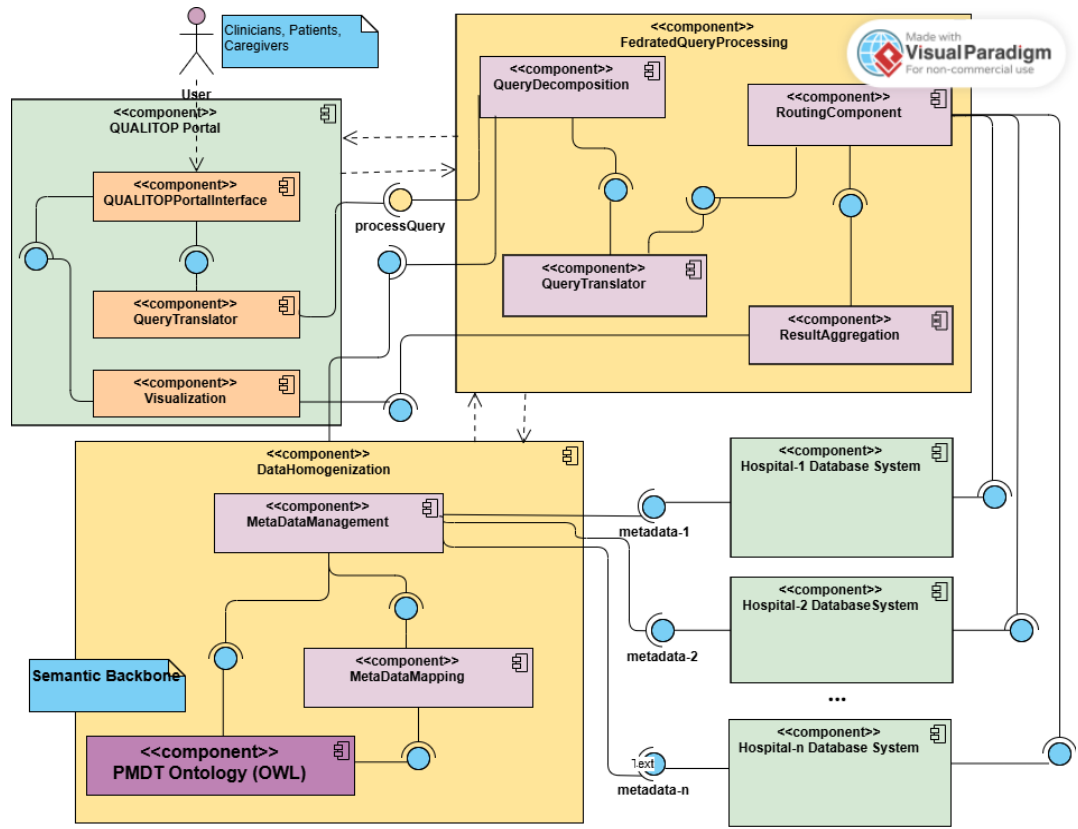

*Fig. 9 UML Component diagram representing the implementation architecture of the PMDT, showing portal, semantic backbone, federated query, and hospital data sources.*

1. **Portal Component**: This serves as the entry point for end-users such as clinicians, nurses, patients, and caregivers. Through a clinical dashboard, users submit both descriptive and analytical queries and get the results visualized. The portal is designed to hide the underlying complexity of federated data access while providing personalized, ontology-driven dashboards. For example, Fig. 10 illustrates a prototype of the QUALITOP portal, where users can enter multimodal patient information (e.g., demographics, lifestyle, treatment details) and receive analytical outputs such as predicted quality-of-life indicators across physical, emotional, and social dimensions. This exemplifies how the PMDT ontology supports not only semantic integration but also downstream predictive and prescriptive analytics.

2. **Data Homogenization Component**: At the core of the architecture, the PMDT ontology (implemented in OWL 2.0) provides the semantic backbone that unifies heterogeneous hospital data sources. Local database schemas are aligned to the ontology through a metadata management and mapping layer, ensuring that multimodal data, ranging from clinical records and imaging to genomics, lifestyle, and quality-of-life questionnaires, can be consistently discovered and queried across institutions. By resolving not only syntactic but also semantic heterogeneity, this component enables federated data access and advanced analytics while preserving the autonomy of local hospital systems.

3. **Federated Query Processing Component**: This component operationalizes ontology-driven queries across distributed hospital databases. User queries submitted through the portal are first expressed in SPARQL and then decomposed into sub-queries that correspond to the physical location of the relevant data. The query processing pipeline includes decomposition, translation into local query languages (e.g., SQL), routing to hospital systems, and aggregation of partial results. By leveraging the semantic mappings defined in the Data Homogenization component, the federated query processor enables transparent, ontology-based access to multimodal patient data while ensuring that local hospital autonomy and data privacy are preserved.

4. **Hospital Database Systems**: Each participating hospital maintains full ownership and control over its local database systems, which may store diverse types of patient data (e.g., EHRs, imaging, QoL surveys, genomic profiles). These systems expose only metadata and query interfaces to the federation, never raw datasets, thereby preserving compliance with GDPR and institutional privacy policies. Through the metadata mappings established in the Data Homogenization layer, hospital databases can be queried in a uniform way, regardless of schema heterogeneity. This federated, privacy-preserving architecture ensures that the PMDT can integrate knowledge across institutions without requiring data centralization.

The architecture highlights how the PMDT ontology functions as the semantic backbone of the ecosystem: it mediates between heterogeneous databases, supports federated reasoning, and provides meaningful results back to end-users through the portal. Its modular design has been validated iteratively in the QUALITOP project and provides the foundation for advanced analytical capabilities. The next section (5.2) details the implementation of the ontology in OWL, which enables semantic interoperability and reasoning over multimodal data, while Section 6 evaluates the overall ecosystem, including expert validation of the portal and federated query mechanisms.

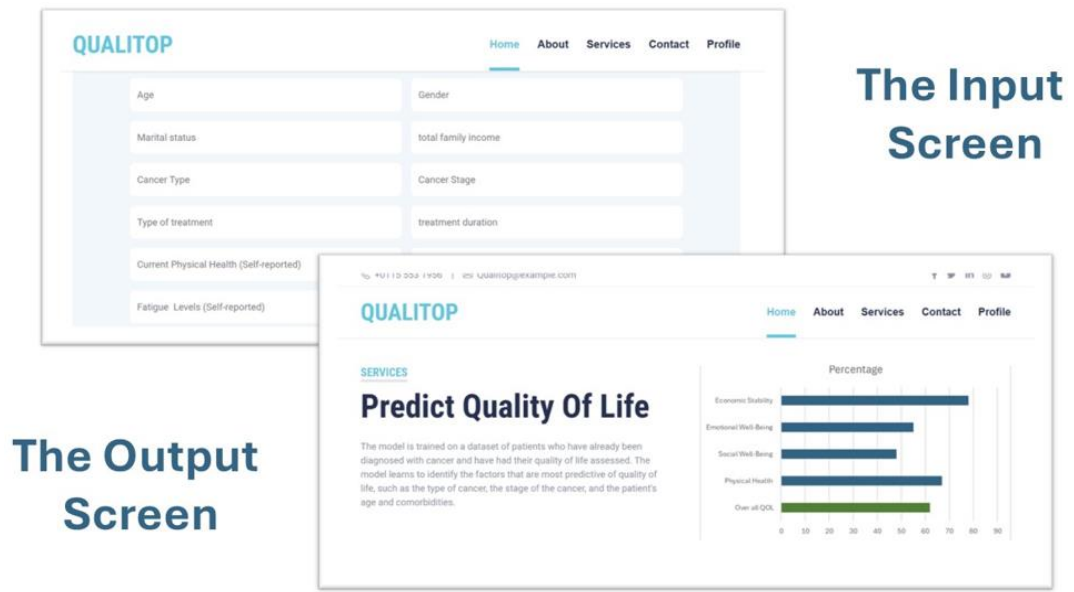

*Fig. 10 QUALITOP portal prototype: user input of multimodal patient data (left) and output of predicted quality-of-life indicators (right), powered by the PMDT ontology.*

## 5.2 PMDT Ontology in OWL

The PMDT ontology was implemented in *OWL 2.0* using the Protégé environment, following the conceptual design introduced in Section 4. Each blueprint of the ontology (Patient, Disease and Diagnosis, Treatment and Follow-up, Adverse Events, Trajectories, Safety, and Medical Pathways) is encoded as an OWL class hierarchy, enriched with object properties, datatype properties, and logical restrictions. This formalization enables automated reasoning, consistency checking, and SPARQL-based querying.

The ontology reuses and aligns with established vocabularies where appropriate. Temporal aspects of trajectories are represented using the *W3C OWL-Time ontology*, ensuring interoperability for modeling *TimeInstant* and *TimeInterval*. For diseases, the ontology is linked to the *Disease Ontology (DO)* to avoid duplication of established taxonomies, while the representation of adverse events integrates CTCAE-based severity grades. By aligning internal classes with these external standards including conceptual mappings to HL7 FHIR and OMOP CDM, with full implementation planned in future work, PMDT ensures extensibility and compatibility with broader biomedical ontologies.

At the implementation level, the ontology serves as the *global schema* mediating heterogeneous data sources (see Section 5.1). Local hospital schemas are semantically mapped to PMDT classes and properties, enabling federated queries to be uniformly expressed in SPARQL. This mapping layer also allows multimodal patient data (e.g., clinical records, genomic markers, imaging, lifestyle, quality-of-life questionnaires) to be retrieved and reasoned over in a consistent way across institutions.

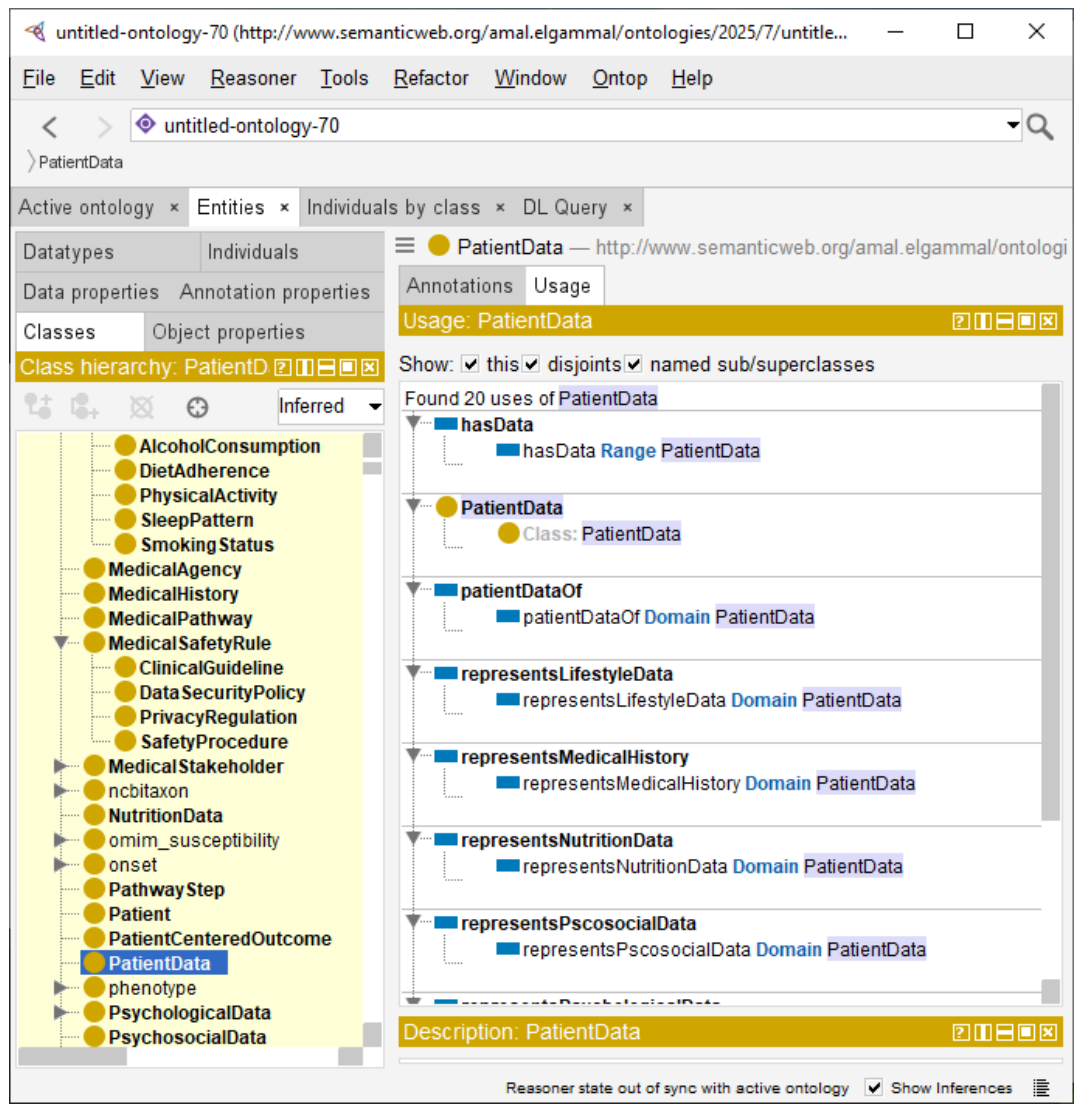

*Fig. 11 Implementation of the PMDT ontology in OWL 2.0 using Protégé, showing the PatientData hierarchy and its multimodal dimensions.*

Fig. 11 and Fig. 12 show screenshots of the ontology implementation in Protégé, illustrating the OWL representation of the *PatientData* and *Treatment* hierarchies, respectively. The PatientData hierarchy formalizes multimodality by linking lifestyle, nutrition, psychosocial, psychological, and medical history elements, ensuring that diverse patient data streams can be semantically integrated and queried consistently. The Treatment hierarchy captures interventions such as cancer therapy, medication, surgery, and follow-up processes, while linking them to outcomes, adverse events, and temporal aspects. Together, these hierarchies demonstrate how the PMDT ontology operationalizes both multimodal evidence and treatment processes at the implementation level. The use of OWL 2.0 supports descriptive analytics (e.g., retrieving patients with a given comorbidity or treatment history) and provides a foundation for predictive and prescriptive reasoning when combined with AI/ML techniques. The feasibility and validation of these capabilities are discussed in Section 6.

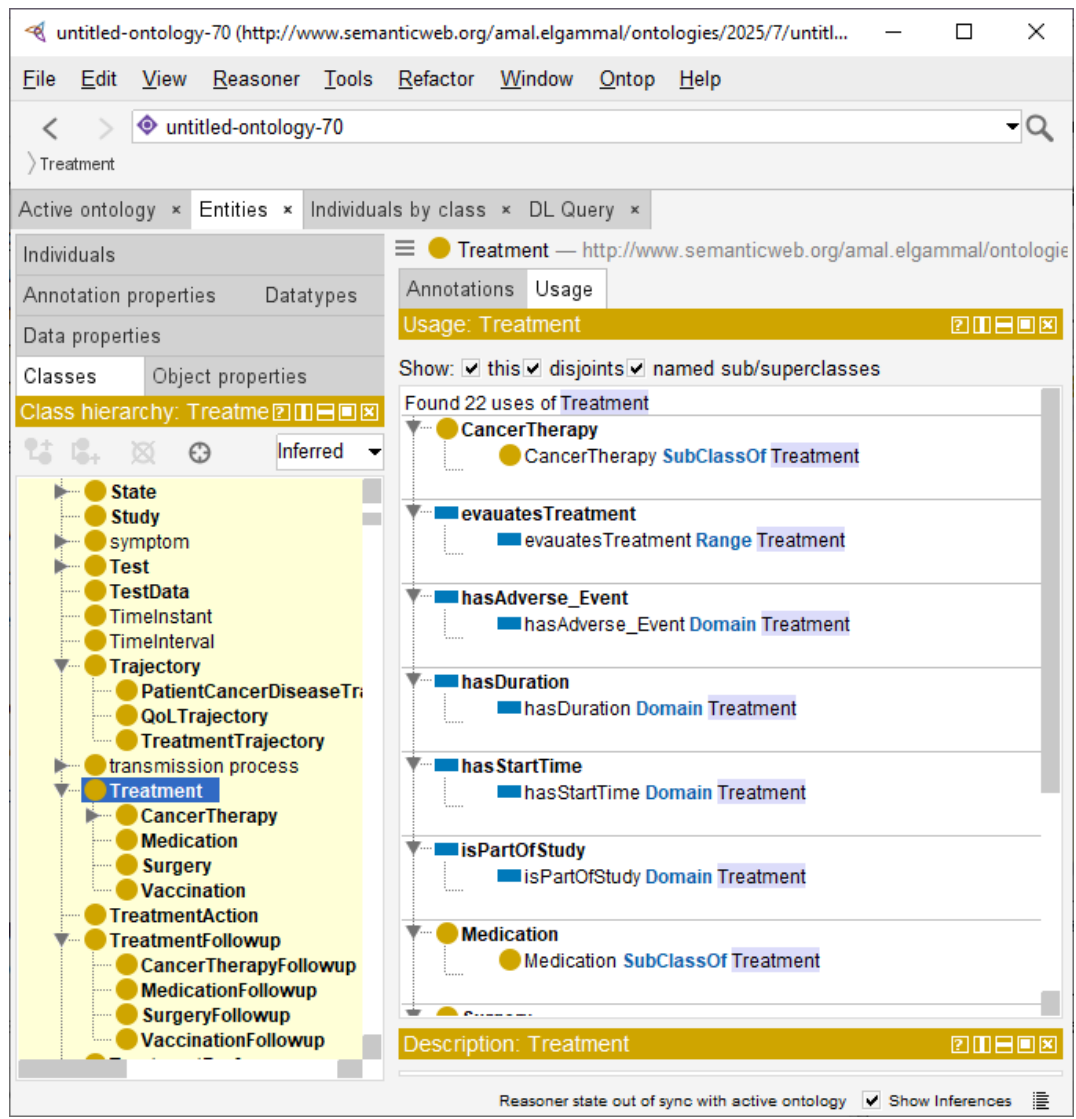

*Fig. 12 PMDT ontology in OWL 2.0 (Protégé), showing the Treatment hierarchy with cancer therapies, interventions, and follow-ups.*

# 6 Validation and Evaluation

The validation of the Patient Medical Digital Twin (PMDT) combined evidence from three complementary approaches: (i) a multicenter pilot study within the EU H2020 QUALITOP project, (ii) expert feedback through workshops, interviews, and questionnaires, and (iii) evaluation against predefined criteria covering ontology coverage, analytical capabilities, usability, and compliance. The following subsections present the pilot study setup (Section 6.1), expert validation (Section 6.2), evaluation criteria (Section 6.3), and the results and discussion (Sections 6.4–6.5).

## 6.1 Pilot Study Setup

The validation of the PMDT ontology was carried out in the context of the EU H2020 QUALITOP project, through an international, observational multicenter cohort study [35] conducted in France, the Netherlands, Portugal, and Spain. The study[2] included adult patients treated with CAR-T cell therapy or immune checkpoint inhibitors, two immunotherapies known for their clinical complexity, the frequent occurrence of immune-related adverse events (irAEs), and their impact on quality of life (QoL). This real-world clinical setting provided a rich environment to evaluate the PMDT ontology, particularly its ability to represent multimodal patient data, treatment trajectories, adverse events, and QoL outcomes. Patients were monitored for up to 18 months post-treatment initiation, with data collected at baseline and follow-up visits (3, 6, 12, and 18 months). The study captured diverse modalities including clinical records, laboratory data, psychosocial and behavioral factors, registries and administrative databases, and patient-reported QoL questionnaires. This heterogeneity closely aligned with the *PatientData blueprint* of the PMDT ontology, which is designed to integrate multimodal evidence across time.

The pilot study also served to validate the competency questions formulated in Section 4.2. For example, longitudinal monitoring of QoL and clinical outcomes directly supported *trajectory-related queries* (e.g., "Compare a patient's QoL

---

[2] The QUALITOP cohort study has been registered as an observational study at www.clinicaltrials.gov. Trial registration number NCT05626764, 2022-11-25

before and 3 months after CAR-T infusion"), while the recording of immune-related adverse events enabled queries from the *Medical Safety and Adverse Events views* (e.g., "Which irAEs were reported after immunotherapy, and how were they managed?"). Similarly, the multimodal integration of lifestyle, psychosocial, and genomic data provided evidence for *Main View questions* (e.g., "How can lifestyle and psychosocial data be integrated with diagnostic evidence to personalize treatment and follow-up?").

It is important to note that while the ontology was designed to support a wide range of competency questions, only a representative subset was implemented as a *proof-of-concept* within QUALITOP. Specifically, queries addressing multimodal integration, adverse event monitoring, and longitudinal QoL assessment were prioritized, reflecting the most pressing clinical needs of the pilot. This selective focus ensured that evaluation remained clinically relevant and feasible, while leaving the broader set of queries available for future extensions.

By grounding competency questions in real-world, heterogeneous patient data, the pilot study demonstrated that the PMDT ontology is both clinically relevant and operationally feasible, capable of supporting not only *descriptive queries* but also laying the foundation for *predictive and prescriptive analytics* in oncology care.

## 6.2 Expert feedback Process: Interviews and Questionnaire

The development and validation of the PMDT ontology followed an iterative, expert-driven methodology, involving continuous feedback from oncologists, immunologists, healthcare IT specialists, and patients' representatives participating in the QUALITOP project. Feedback was collected through structured questionnaires, semi-structured interviews, and regular workshops across the project's duration.

This process ensured that the ontology remained clinically relevant and aligned with user needs. Specifically, experts highlighted four key points:

- **Validation of scope and coverage** against real-world immunotherapy cases, confirming the ontology's ability to represent multimodal data, treatment trajectories, adverse events, and QoL outcomes.

- **Refinement of competency questions** introduced in Section 4.2, ensuring that they corresponded to meaningful clinical queries (e.g., management of irAEs, impact of comorbidities, longitudinal QoL monitoring).
- **Confirmation of usability requirements** for query execution through the portal interface, stressing the importance of descriptive, predictive, and prescriptive queries in clinical practice.
- **Identification of data protection and compliance issues**, particularly around GDPR, emphasizing the need for privacy-by-design measures and institutional Data Protection Officer involvement.

As one patient representative remarked, *"patients care deeply about whether projects like this have been reviewed by ethics committees and validated for GDPR compliance , it is essential for building trust."*

Overall, the expert feedback validated the PMDT ontology as a semantically robust and clinically usable knowledge model, while also identifying areas for future improvement, such as piloting the platform with real patients in controlled hospital environments, enhancing regulatory alignment, and further tailoring user interfaces for non-technical stakeholders.

### *6.2.1 Questionnaire-based Evaluation*

To complement the expert workshops described above, a structured questionnaire was designed to gather formative feedback on the implemented PMDT prototype. The evaluation aimed not to provide statistical generalization, but to assess the perceived utility, usability, and domain relevance of the ontology-driven system after guided use.

**Evaluation protocol:** Participants included medical experts (oncologists, immunologists, and healthcare IT specialists) and patients' representatives engaged in the QUALITOP project, together with an external oncology research group in Egypt. Before completing the questionnaire, all respondents participated in a 30-minute demonstration and hands-on walkthrough of the QUALITOP portal. They were asked to perform example tasks aligned with the ontology's competency questions, such as:

- submitting a QoL prediction query for a melanoma patient under immunotherapy,

- exploring adverse events and their severity grades, and
- navigating multimodal patient data (clinical history, lifestyle, psychosocial).

The questionnaire was then administered immediately after interaction, ensuring that responses reflected actual system use rather than abstract expectations.

**Questionnaire structure:** The instrument was organized into four sections: (i) data homogenization and semantic integration, (ii) analytical and reasoning capabilities, (iii) usability and interaction design, and (iv) technical aspects and areas for improvement. Each item was rated on a 5-point Likert scale (1 = strongly disagree, 5 = strongly agree). This structure distinguished between the system's technical foundations and its clinical relevance, addressing concerns of mixing generic and specific issues.

**Participants:** Twelve responses were collected (75% response rate) from 16 invited participants: four QUALITOP medical partners, two patient representatives, and six oncology researchers from Baheya Hospital (https://baheya.org/), Egypt. While the sample size is modest, it provided diverse perspectives from both European and external medical contexts.

**Results overview:** Table 2 summarizes descriptive statistics. The strongest agreement was observed for the need to address heterogeneity of patient-related data (mean 4.8, 100% positive responses), and for the portal's role-based access and privacy measures (mean 4.7, 91.6% positive responses). Analytical capabilities were also highly rated (mean 4.3–4.6), particularly the ability to support decision-making without requiring sensitive data transfer. Usability aspects were evaluated positively overall (mean 4.0–4.2), though some respondents emphasized the need for further simplification of the user interface.

As one participant remarked, *"The platform's ability to unify data sources while preserving privacy is a major step forward, but the interface must remain simple enough for everyday clinical use."*

**Interpretation:** Although exploratory in scope, the questionnaire confirmed that the PMDT ontology and portal prototype are seen as both technically feasible and clinically relevant. The results reinforce the ontology's competency questions and highlight future priorities: piloting with real patients, refining UI design, and extending regulatory validation.

Table 2 Summary of questionnaire responses (n=12), grouped into domain-relevance and system-oriented categories.

| Section | Question Focus | Example Items | Mean (SD) | % Agree (4–5) |
|---|---|---|---|---|
| **Domain Relevance** | Data heterogeneity & integration | "The heterogeneity, diversity, and fragmentation of patient data represent a real challenge in gaining a holistic view." | 4.8 (0.5) | 100% |
| | Clinical value of PMDT | "The platform helps integrate multimodal patient data (clinical, QoL, lifestyle) to provide a holistic view." | 4.2 (0.9) | 83% |
| | Analytical reasoning for decision support | "The analytics and learning capabilities increase my insights and support decision-making." | 4.3 (1.4) | 75% |
| **System-Oriented Aspects** | Usability & interaction | "The platform interface is user-friendly and easy to navigate." | 4.0 (0.8) | 79% |
| | Security & privacy | "Role-based access and GDPR-compliant measures increase my trust in the platform." | 4.7 (0.7) | 92% |
| | Technical extensibility | "The platform's architecture can be applied across other healthcare domains." | 4.5 (1.0) | |

## 6.3 Evaluation Criteria

The evaluation of the PMDT ontology and its implementation followed a dual perspective: *(i) domain relevance*, assessing whether the ontology adequately represents clinically meaningful concepts and supports the competency questions formalized in Section 4, and *(ii) system-oriented usability*, evaluating whether the implemented ecosystem enables experts to query, interpret, and act upon multimodal data in practice.

Four high-level criteria were defined:

1. **Coverage and Accuracy of Ontology Models (Domain Relevance):** Experts assessed whether the PMDT sufficiently covers the key domains of oncology care (disease and diagnosis, treatment and follow-up, adverse events, trajectories, safety, and pathways). Particular attention was paid to whether the ontology supports clinically meaningful queries, such as those formalized as competency questions in Section 4. To remain realistic within the QUALITOP pilot, only a representative subset of these queries (focused on multimodal integration, irAE monitoring, and QoL trajectories) was implemented as a proof-of-concept.

2. **Analytical Capabilities:** The ability of the ontology-driven system to support different levels of analytics was assessed. This included descriptive queries (e.g., retrieving patients with a given comorbidity), predictive queries (e.g., forecasting QoL outcomes), and prescriptive queries (e.g., recommending treatment adjustments). The evaluation focused on whether implemented queries could combine multimodal data (e.g., treatment trajectory, lifestyle, and QoL) to yield actionable insights.

3. **Usability and User Experience:** The QUALITOP portal was evaluated for ease of use, clarity of visualizations, and perceived utility in daily workflows. Clinicians, patients' representatives, and IT specialists provided structured feedback on whether the portal effectively abstracts the underlying complexity of federated queries and delivers results in a clinically interpretable form.

4. **Compliance and Trustworthiness:** Experts evaluated whether the federated architecture and portal design reflected *privacy-by-design principles*, GDPR compliance, and ethical acceptability. These factors were considered critical to building trust among both healthcare professionals and patients in adopting ontology-driven digital twins.

Together, these criteria ensured that the evaluation balanced *semantic robustness* (ontology coverage and reasoning support) with *practical applicability* (analytics, usability, and compliance).

## 6.4 Results

The evaluation produced complementary findings along the four defined criteria:

1. **Coverage and Accuracy of Ontology Models:** Domain experts confirmed that the PMDT ontology adequately captured the major conceptual domains of oncology care, including *disease and diagnosis, treatment and follow-up, adverse events, trajectories, safety, and pathways*. The ontology was able to represent clinically relevant concepts such as immune-related adverse events (irAEs), treatment trajectories, and longitudinal quality-of-life (QoL) outcomes. Importantly, while the ontology supports a wide range of competency questions defined in Section 4, only a *subset of representative queries* was implemented during the QUALITOP pilot as proof-of-concept, with a focus on QoL monitoring, multimodal integration, and irAE tracking. Experts highlighted this as a pragmatic step that demonstrated feasibility without overstating maturity.

2. **Analytical Capabilities:** The system successfully executed *descriptive queries* (e.g., retrieving patients with comorbidities or extracting recorded irAEs) and supported initial *predictive tasks* (e.g., forecasting QoL at 3–12 months post-treatment using integrated multimodal data). While full prescriptive analytics (e.g., treatment adjustment recommendations) were not implemented, experts agreed that the ontology design provides a strong foundation for extending to prescriptive use cases in future work. The ability to combine heterogeneous evidence (clinical, psychosocial, and patient-reported data) in a single query was viewed as a major advance over current fragmented systems.

3. **Usability and User Experience:** Feedback on the QUALITOP portal prototype was generally positive. Clinicians emphasized the value of having *ontology-driven dashboards* that hid the complexity of federated queries and returned results in an interpretable form. Patients' representatives appreciated the incorporation of QoL indicators, which they saw as critical for shared decision-making. However, experts suggested refinements to improve *visualization clarity* (e.g., better trend charts for longitudinal QoL trajectories) and to further simplify *query formulation* for non-technical users.

4. **Compliance and Trustworthiness:** The federated architecture and metadata-driven approach were recognized as strong enablers of *privacy-preserving data sharing*. The explicit encoding of GDPR and safety rules in the ontology (see Section 4.2.2, Medical Safety view) was positively evaluated as a *unique contribution*. Patient representatives stressed the importance of ethics oversight and GDPR validation by institutional Data

> Protection Officers, echoing their concern that "trust in digital health systems depends as much on governance as on functionality."

Overall, the evaluation validated the PMDT ontology as a *semantically robust and clinically meaningful model*, with a technically feasible implementation in the QUALITOP pilot. At the same time, it highlighted areas for refinement: extending prescriptive analytics, enhancing visualization and usability, and continuing regulatory alignment. These results demonstrate both the *current readiness* of the PMDT as a knowledge backbone and its *future potential* for predictive and prescriptive decision support in oncology care.

## 6.5 Discussion

The evaluation results highlight both the strengths and limitations of the current PMDT implementation. Conceptually, the ontology design presented in Section 4 , structured around Blueprints and specialized views (Main, Treatment and Follow-up, Trajectory, Safety, Pathways, Adverse Events), proved effective in covering the essential dimensions of oncology care. Experts confirmed that the ontology supports clinically meaningful queries and provides a structured foundation for reasoning across multimodal data sources. This validates the design choice of adopting a *blueprint-based, modular ontology*, which enables extensibility and alignment with external standards such as OWL-Time and the Disease Ontology.

From an implementation perspective (Section 5), the QUALITOP pilot demonstrated the feasibility of embedding the ontology into a federated, privacy-preserving architecture. The *Portal* provided a usable entry point for clinicians and patients' representatives, while the *Data Homogenization* and *Federated Query Processing* components ensured semantic interoperability across distributed hospital databases. Together, these elements operationalized the conceptual vision described in Section 3, where multimodal integration, longitudinal reasoning, and privacy-preserving analytics were identified as key requirements.

In terms of analytics, the PMDT has shown immediate value for *descriptive and predictive queries*, such as tracking immune-related adverse events or forecasting short-term quality-of-life outcomes. However, the evaluation also underlined that *prescriptive capabilities* (e.g., recommending treatment adjustments or preventive strategies) remain at a conceptual level. Realizing these use cases will require extending the ontology with more fine-grained rules, validated predictive models, and iterative testing with larger patient cohorts.

An important lesson from the validation is the central role of *trust and usability*. Experts emphasized that even the most sophisticated ontology and federated architecture will only be adopted if the system is intuitive to use and demonstrably compliant with regulatory frameworks such as GDPR. Here, the explicit encoding of safety rules and privacy regulations in the ontology (Medical Safety view) was seen as a unique and promising step. At the same time, clinicians requested clearer visualizations of longitudinal data, while patient representatives stressed the need for continuous ethics and governance oversight.

Finally, the evaluation confirmed the value of the persona-based approach (Section 3.3) for grounding technical requirements in realistic clinical contexts. The trajectories of Elena (early-stage melanoma with diabetes), Markus (advanced melanoma under immunotherapy), and Aisha (genomic risk for melanoma) resonated with experts as representative scenarios, and the competency questions derived from them provided concrete validation points for the ontology. This suggests that personas are not only useful as illustrative devices in research papers, but also as practical anchors for aligning ontology development with clinical needs. In summary, the QUALITOP pilot demonstrated that the PMDT ontology is both *semantically robust and clinically meaningful*, while also identifying clear directions for improvement: (i) extending prescriptive analytics, (ii) refining portal usability and visualization, and (iii) strengthening regulatory integration. These findings underscore the potential of the PMDT to evolve from a proof-of-concept knowledge backbone into a *next-generation decision-support framework* for personalized, predictive, and preventive oncology care.

## 8 Conclusion & Future Work

This paper introduced the Patient Medical Digital Twin (PMDT), a formally modelled, ontology-driven framework that unifies multimodal health data (clinical, psychosocial, behavioral, and genomic) into a semantically interoperable, extensible, and privacy-preserving ecosystem. Unlike organ- or disease-specific twins, the PMDT offers a holistic, continuously updated representation of the patient, enabling proactive prevention as well as precision care.

The ontology, developed in OWL 2.0 and structured around modular Blueprints, provides the semantic backbone for representing disease and diagnosis, treatments and follow-ups, trajectories, safety, pathways, and adverse events. Its embedding into a federated implementation architecture (Section 5) demonstrates how the PMDT can mediate heterogeneous hospital databases, support ontology-driven queries, and deliver actionable insights through a user-facing portal.

Validation in the QUALITOP pilot study (Section 6) confirmed the clinical relevance and feasibility of the PMDT. Experts validated the ontology's ability to represent multimodal patient data, support competency questions derived from realistic personas, and enable descriptive and predictive analytics in practice. At the same time, evaluation highlighted areas for future improvement, particularly in prescriptive decision support, usability of visual interfaces, and regulatory alignment.

Crucially, the modular and ontology-driven design ensures that the PMDT is not static: it is extensible to incorporate additional modalities such as biosensor and imaging data, preparing the ground for broader clinical deployment and more advanced analytics.

Future work will extend the PMDT in four key directions:

1. **Prescriptive and adaptive analytics**: moving beyond descriptive queries to recommend personalized interventions.
2. **Secure and transparent health data exchange**: validating the implemented provenance constructs (*DataSource*, *ConsentStatement*, *AccessPolicy*) and exploring blockchain and advanced governance mechanisms to strengthen trust and compliance.
3. **Expanded data modalities**: incorporating biosensor and imaging data streams to enrich multimodality and enable continuous, real-time patient monitoring within the PMDT framework.
4. **Broader clinical evaluation**: expanding validation beyond melanoma to other cancer types, including colorectal cancer screening in the ONCOSCREEN project (https://cordis.europa.eu/project/id/101097036).

By combining formal semantic modeling with a modular, privacy-preserving architecture and grounding it in real-world validation, the PMDT lays the foundation for the next generation of intelligent, clinically validated medical digital twins. Ultimately, the PMDT vision is to transform chronic care from reactive treatment toward proactive, continuously optimized, and universally accessible health management.

**Author Contribution** *Amal Elgammal* contributed to the conceptualization and led the detailed modeling of the PMDT ontology, implemented it in OWL, and coordinated the iterative requirements engineering and validation with medical partners. She also contributed substantially to the manuscript writing and integration. *Bernd J. Krämer* contributed to conceptualization, provided guidance on methodological aspects, and reviewed and refined the manuscript. *Michael P. Papazoglou* conceived the overall idea of the digital health ecosystem, led to the conceptualization of the PMDT blueprints and high-level architecture, and authored major parts of the manuscript. *Mira Raheem* implemented the digital platform components, including the collaborative portal and federated analytics capabilities, and contributed to the implementation section of the manuscript.

**Funding** This research is partially funded by the EC Horizon 2020 project QUALITOP, under contract number H2020 - SC1-DTH-01-2019 – 875171.

**Availability of Data and Materials** The data sets used are not publicly accessible for data protection reasons under data transfer agreements signed with Medical Partners in the QUALITOP Project.

**Clinical Trial Number** Not applicable.

**Ethical Approval** Not applicable.

**Competing Interests** The authors declare no financial or non-financial competing interests.

**Acknowledgements**

We thank Hospital Clinic Barcelona (IDIBAPS), Hospices Civils de Lyon (HCL), University Medical Center Groningen (UMCG), and Instituto Português de Oncologia de Lisboa (IPOL) for their contributions to the pilot study and their continuous support in the evaluation and validation of this work. We also extend our gratitude to Dr. Roxana Albu, Chief Scientific Officer at the Association of European Cancer Leagues (ECL), and Dr. Menia Koukougianni, Fellow of the European Patients' Academy and Baheya Hospital (Egypt), for their valuable feedback in assessing the clinical relevance and patient-centered aspects of the PMDT.